\documentclass{sig-alternate}
\usepackage[noadjust]{cite}
\usepackage{graphicx}
\usepackage{subfigure}

\newcommand{\beq}{\begin{equation}}
\newcommand{\eeq}{\end{equation}}
\def\bearn{\begin{eqnarray*}}
\def\eearn{\end{eqnarray*}}
\def\barr{\begin{array}}
\def\earr{\end{array}}

\def\bt{BitTorrent }
\def\btns{BitTorrent} 
\def\p2p{peer-to-peer}

\begin{document}
\conferenceinfo{IMC'06,} {October 25--27, 2006, Rio de Janeiro, Brazil.}
\CopyrightYear{2006}
\crdata{1-59593-561-4/06/0010} 
\toappear{© ACM, 2006. This is the author's version of the work. It
  is posted here by permission of ACM for your personal use. Not for
  redistribution. The definitive version was published in
  Proc. of IMC'06, October 25--27, 2006, Rio de Janeiro, Brazil.} 

\title{Rarest First and Choke Algorithms Are Enough}

\numberofauthors{2}
\author{
\alignauthor Arnaud Legout\\
\affaddr{I.N.R.I.A.}\\
\affaddr{Sophia Antipolis}
\affaddr{France}
\email{arnaud.legout@sophia.inria.fr}
\alignauthor G. Urvoy-Keller and P. Michiardi\\
\affaddr{Institut Eurecom}\\
\affaddr{Sophia Antipolis}
\affaddr{France}
\email{{\sffamily\large\bfseries\{}Guillaume.Urvoy,Pietro.Michiardi{\sffamily\large\bfseries\}}@eurecom.fr}
}

\maketitle

\sloppy
\begin{abstract}
  The performance of peer-to-peer file replication comes from its
  piece and peer selection strategies. Two such strategies have been
  introduced by the BitTorrent protocol: the rarest first and choke
  algorithms. Whereas it is commonly admitted that BitTorrent performs
  well, recent studies have proposed the replacement of the rarest
  first and choke algorithms in order to improve efficiency and
  fairness. In this paper, we use results from real experiments to
  advocate that the replacement of the rarest first and choke
  algorithms cannot be justified in the context of peer-to-peer file
  replication in the Internet.

  We instrumented a BitTorrent client and ran experiments on real
  torrents with different characteristics. Our experimental evaluation
  is peer oriented, instead of tracker oriented, which allows us to
  get detailed information on all exchanged messages and protocol
  events.  We go beyond the mere observation of the good efficiency of
  both algorithms. We show that the rarest first algorithm guarantees
  close to ideal diversity of the pieces among peers. In
  particular, on our experiments,  replacing the rarest first
  algorithm with source or network coding solutions cannot be
  justified. We also show that the choke algorithm in its latest
  version fosters reciprocation and is robust to free riders. In
  particular, the choke algorithm is fair and its replacement with a
  bit level tit-for-tat solution is not appropriate. Finally, we
  identify new areas of improvements for efficient peer-to-peer file
  replication protocols.
\end{abstract}

\vspace{1mm}
\noindent
{\bf Categories and Subject Descriptors:} C.2.2
{[Computer-Communication Networks]}: {Network Protocols}; C.2.4
{[Computer-Communication Networks]}: {Distributed Systems}

\vspace{1mm}
\noindent
{\bf General Terms:} Measurement, Algorithms, Performance

\vspace{1mm}

\section{Introduction}
\label{sec:introduction}
In a few years, \p2p file sharing has become the most popular
application in the Internet \cite{karagiannis04_imc, karagiannis04}.
Efficient content localization and replication are the main reasons
for this success. Whereas content localization has attracted
considerable research interest in the last years \cite{stoica01,
  ratnasamy01, chawathe03, gummadi03}, content replication has started
to be the subject of active research only recently.  As an example,
the most popular \p2p file sharing networks \cite{slyck} eDonkey2K,
FastTrack, Gnutella, Overnet focus on content localization. The only
widely used \cite{karagiannis04_imc, karagiannis04, parker04} \p2p
file sharing application focusing on content replication is \bt
\cite{cohen03}.

Yang et al. \cite{yang04} studied the problem of efficient content
replication in a \p2p network. They showed that the capacity of the
network to serve content grows exponentially with time in the case
of a flash crowd, and that a key improvement on \p2p file replication
is to split the content into several pieces. Qiu et al. \cite{qiu04}
proposed a refined model of \bt and showed its high efficiency. In
summary, these
studies show that a \p2p architecture for file replication is a major
improvement compared to a client server architecture, whose capacity
of service does not scale with the number of peers.

However, both studies assume global knowledge, which is not
realistic. Indeed, they assume that each peer knows all the other
peers.
As a consequence, the results obtained with
this assumption can be considered as the optimal case. In real
implementations, there is no global knowledge. The challenge is then
to design a \p2p protocol that achieves a level of efficiency close to
the one achieved in the case of global knowledge.

Piece and peer selection strategies are the two keys of efficient \p2p
content replication. Indeed, in a \p2p system, the content is split
into several pieces, and each peer acts as a client and a server.
Therefore, each peer can receive and give any piece to any other peer.
An efficient piece selection strategy should guarantee that each peer
can always find an interesting piece from any other peer. The rationale
is to offer the largest choice of peers to the peer selection
strategy. An efficient peer selection strategy should maximize the
capacity of service of the system. In particular, it
should employ selection criteria based, e.g., on upload and download
capacity, and should not be biased by the lack of available pieces in
some peers.

The rarest first algorithm is a piece selection strategy that consists
of selecting the rarest pieces first. This simple strategy used by \bt
performs better than random piece selection strategies
\cite{bharambeApr06,felber04}. However, Gkantsidis et al.
\cite{gkantsiInf05} argued based on simulations that the rarest first
algorithm may lead to the scarcity of some pieces of content and
proposed a solution based on network coding. Whereas this solution is
elegant and has raised a lot of interest, it leads to several complex
deployment issues such as security and computational cost.  Other
solutions based on source coding \cite{kosticUsenix05} have also been
proposed to solve the claimed deficiencies of the rarest first
algorithm.

The choke algorithm is the peer selection strategy of \btns. This
strategy is based on the reciprocation of upload and download speeds.
Several studies \cite{GuoIMC05,JunSigWork05,Ganesan05,bharambeApr06}
discussed the fairness issues of the choke algorithm. In
particular, they argued that the choke algorithm is unfair and
favors free riders, i.e., peers that do not contribute. Solutions based
on a bit level tit-for-tat have been proposed to address the choke
algorithm's fairness problem.

In this paper, we perform an experimental evaluation of the piece and
peer selection strategies as implemented in \btns. Specifically, we
have instrumented a client and run extensive experiments on several
torrents with different characteristics in order to evaluate the
properties of the rarest first and choke algorithms. While we have not
examined all possible cases, we argue that we have covered a
representative set of today real torrents.

Our main conclusions on real torrents are the following.
\begin{itemize}
\item The rarest first algorithm guarantees a high diversity of the
  pieces. In particular, it prevents the reappearance of rare pieces and
  of the last pieces problem.
\item We have found that torrents in a startup phase can have low
  piece diversity. The duration of this phase depends only on the
  upload capacity of the source of the content. In particular, the
  rarest first algorithm is not responsible for the low piece
  diversity during this phase.
\item The fairness achieved with a bit level tit-for-tat strategy is
  not appropriate in the context of \p2p file replication. We have
  proposed  two new fairness criteria in this context.
\item The choke algorithm is fair, fosters reciprocation, and is
  robust to free riders in its latest version.
\end{itemize}

Our contribution is to go beyond the mere confirmation of the good
performance of \btns. We provide new insights into the role of peer
and piece selection for efficient \p2p file replication.  We show for
the first time that on real torrents, the efficiency of the rarest
first and choke algorithms do no justify their replacement by more
complex solutions. Also, we identify, based on our observations, new
area of improvements: the replication of the first pieces and the
speed of delivery of the first copy of the content. Finally, we
propose two new fairness criteria in the context of \p2p file
replication and we present for the first time results on the new
version of the choke algorithm that fixes fundamental fairness issues.

Our findings significantly differ from previous work \cite{gkantsiInf05, kosticUsenix05,
  GuoIMC05,JunSigWork05,Ganesan05,bharambeApr06}. There are three main
reasons for this divergence. First, we target \p2p file
replication in the Internet. As a consequence, the peers are well
connected without severe network bottlenecks. The problems
identified in the literature with the rarest first algorithm are in
the context of networks with connectivity problems or low capacity
bottlenecks.  Second, we evaluate for the first time the new version
of the choke algorithm. The evaluation of the choke algorithm in the
literature was performed on the old version. We show that the new
version solves the problems identified on the old one.  Finally, we
perform an experimental evaluation on real torrents.  Simulating
\p2p protocols is hard and requires many simplifications. In
particular, all the simulations of \bt we are aware of consider that
each peer only knows few other peers, i.e., each peer has a small peer
set \cite{bharambeApr06, gkantsiInf05}. In the case of real torrents, the peer set
size is much larger.  The consequence is that \bt builds a random
graph, connecting the peers, that has a larger diameter in simulations
than in real torrents.  However, the diameter has a fundamental impact
on the efficiency of the rarest first algorithm.

In this study, we show that in the specific context considered, i.e.,
Internet \p2p file replication, the rarest first and
choke algorithms are good enough. Even if we cannot extend our conclusions
to other \p2p contexts, we believe this paper sheds new light on a
system that uses a large fraction of the Internet bandwidth. 

The rest of the paper is organized as follows. We present the
terminology used throughout this paper in
section~\ref{sec:terminology}. Then, we give a short overview of the
\bt protocol in section~\ref{sec:bt-overview} and a
description of the rarest first and choke algorithms in section~\ref{sec:bt-algor-descr}.
We present our experimental methodology in
section~\ref{sec:exper-meth}, and our detailed results in
section~\ref{sec:experimentation-results}. Related work is discussed
in section~\ref{sec:related-work}. We conclude the paper with a
discussion of the results in section~\ref{sec:discussion}.

\section{Background}
\label{sec:background}
We introduce in this section the terminology used throughout this
paper. Then, we give an overview of the BitTorrent protocol, and we
present the rarest first and choke algorithms.

\subsection{Terminology}
\label{sec:terminology}
The terminology used in the \p2p community and in particular
in the \bt community is not standardized. For the sake of
clarity, we define in this section the terms used throughout this
paper.

\begin{itemize}
\item
\textbf{Pieces and Blocks} 
Files transfered using \bt are split in
  \textit{pieces}, and each piece is split in \textit{blocks}. Blocks
  are the transmission unit on the network, but the protocol only
  accounts for transfered pieces. In particular, partially received
  pieces cannot be served by a peer, only complete pieces can.

\item
\textbf{Interested and Choked} 
We say that peer $A$ is
  \textit{interested} in peer $B$ when peer $B$ has pieces that peer
  $A$ does not have. Conversely, peer $A$ is \textit{not interested}
  in peer $B$ when peer $B$ only has a subset of the pieces of peer
  $A$. We say that peer $A$ \textit{chokes} peer $B$ when peer $A$
  decides not to send data to peer $B$. Conversely, peer $A$
  \textit{unchokes} peer $B$ when peer $A$ decides to send data to
  peer $B$.

\item
\textbf{Peer Set} 
Each peer maintains a list of other peers it knows
  about. We call this list the \textit{peer set}. The notion of peer
  set is also known as neighbor set.

\item
\textbf{Local and Remote Peers}
 We call \textit{local peer} the peer
  with the instrumented \bt client, and \textit{remote peers} the
  peers that are in the peer set of the local peer.

\item
\textbf{Active Peer Set}
 A peer can only send data to a subset of its
  peer set. We call this subset the \textit{active peer set}. The
  choke algorithm (described in section~\ref{sec:choke-algorithm})
  determines the peers being part of the active peer set, i.e., which
  remote peers will be choked and unchoked.  Only peers that are
  unchoked by the local peer and interested in the local peer are part
  of the active peer set.

\item
\textbf{Leecher and Seed} 
A peer has two states: the \textit{leecher
    state}, when it is downloading content, but does not have yet
  all the pieces; the \textit{seed state} when the peer has all the pieces
  of the content. For short, we say that a peer is a
  \textit{leecher} when it is in leecher state and a \textit{seed}
  when it is in seed state.

\item
\textbf{Initial Seed} 
The \textit{initial seed} is the peer that is the
  first source of the content.

\item
\textbf{Rarest First Algorithm} 
The \textit{rarest first algorithm} is the piece
  selection strategy used in \btns. We give a detailed description of
  this algorithm in section~\ref{sec:rarest-first}. The rarest first
  algorithm is also called the local rarest first algorithm. 

\item
\textbf{Choke Algorithm} 
The \textit{choke algorithm} is the peer selection
  strategy used in \btns. We give a detailed description of this
  algorithm in section~\ref{sec:choke-algorithm}.  The choke algorithm
  is also called the tit-for-tat algorithm, or tit-for-tat like
  algorithm.

\item
\textbf{Rare and Available Pieces}
 We call the pieces only present on
  the initial seed \textit{rare pieces}, and we call the pieces already served
  at least once by the initial seed \textit{available pieces.}

\item
\textbf{Rarest Pieces and Rarest Pieces Set}
 The \textit{rarest pieces} are the
  pieces that have the least number of copies in the peer set. In the
  case the least replicated piece in the peer set has $m$ copies, then
  all the pieces with $m$ copies form the \textit{rarest pieces set}. The
  rarest pieces can be rare pieces or available pieces, depending on
  the number of copies of the rarest pieces. 
  
\end{itemize}

\subsection{\bt Overview}
\label{sec:bt-overview}
\bt is a P2P application that capitalizes on the bandwidth of peers to
efficiently replicate contents on a large set of peers. A specificity
of \bt is the notion of \textit{torrent}, which defines a session of
transfer of a single content to a set of peers. Torrents are
independent. In particular, participating in a torrent does not bring
any benefit for the participation to another torrent. A torrent is
alive as long as there is at least one copy of each piece in the
torrent. Peers involved in a torrent cooperate to replicate the file
among each other using \emph{swarming} techniques \cite{rodrigton02}.
In particular, the file is split in pieces of typically 256 kB, and
each piece is split in blocks of 16 kB. Other piece sizes are
possible.

A user joins an existing torrent by downloading a \textit{.torrent}
file usually from a Web server, which contains meta-information on the
file to be downloaded, e.g., the piece size and the SHA-1 hash values
of each piece, and the IP address of the so-called \textit{tracker} of
the torrent.  The tracker is the only centralized component of \btns,
but it is not involved in the actual distribution of the file. It
keeps track of the peers currently involved in the torrent and
collects statistics on the torrent.

When joining a torrent, a new peer asks to the tracker a list of IP
addresses of peers to build its initial peer set. This list typically
consists of 50 peers chosen at random in the list of peers currently
involved in the torrent. The initial peer set will be augmented by
peers connecting directly to this new peer. Such peers are aware of
the new peer by receiving its IP address from the tracker.  Each peer
reports its state to the tracker every 30 minutes in steady-state
regime, or when disconnecting from the torrent, indicating each time
the amount of bytes it has uploaded and downloaded since it joined the
torrent.  A torrent can thus be viewed as a collection of
interconnected peer sets. If ever the peer set size of a peer falls
below a predefined threshold, typically 20 peers, this peer will
contact the tracker again to obtain a new list of IP addresses of
peers.  By default, the maximum peer set size is 80. Moreover, a peer
should not exceed a threshold of 40 initiated connections among the 80
at each time. As a consequence, the 40 remaining connections should be
initiated by remote peers. This policy guarantees a good
interconnection among the peer sets in the torrent.

Each peer knows the distribution of the pieces for each peer in its
peer set. The
consistency of this information is guaranteed by the exchange of
messages \cite{btwikispec}.
The exchange of pieces among peers is governed by two core algorithms:
the rarest first and the choke algorithms. These algorithms are further
detailed in section~\ref{sec:bt-algor-descr}. 

\subsection{\bt Piece and Peer Selection Strategies}
\label{sec:bt-algor-descr}
We focus here on the two core algorithms of \btns: the rarest first
and choke algorithms. We do not give
all the details of these algorithms, but explain the main ideas
behind them.

\subsubsection{Rarest First Algorithm}
\label{sec:rarest-first}
The rarest first algorithm works as follows. Each peer maintains
a list of the number of copies of each piece in its peer set. It uses
this information to define a rarest pieces set. Let $m$ be the number
of copies of the rarest piece, then the index of each piece with $m$
copies in the peer set is added to the rarest pieces set. The rarest
pieces set of a peer is
updated each time a copy of a piece is added to or removed from its peer
set. Each peer selects the next piece to download at random in its
rarest pieces set.

The behavior of the rarest first algorithm can be modified by three
additional policies. First, if a peer has downloaded strictly less
than 4 pieces, it chooses randomly the next piece to be requested. This is
called the \textit{random first policy}. Once it has downloaded at
least 4 pieces, it switches to the rarest first algorithm.  The aim of
the random first policy is to permit a peer to download its first
pieces faster than with the rarest first policy, as it is important to
have some pieces to reciprocate for the choke algorithm. Indeed, a
piece chosen at random is likely to be more replicated than the rarest
pieces, thus its download time will be on average shorter.

Second, \bt also applies a
\textit{strict priority policy}, which is at the block level. When at least one
block of a piece has been requested, the other blocks of the same
piece are requested with the highest priority. The aim of the
strict priority policy is to complete the download of a piece as fast
as possible. As only complete pieces can be sent, it is important to
minimize the number of partially received pieces.

Finally, the last policy is the \textit{end game mode} \cite{cohen03}.
This mode starts once a peer has requested all blocks, i.e., all blocks
have either been already received or requested. While in this mode, the peer
requests all blocks not yet received to all the peers in its peer set
that have the corresponding blocks.  Each time a block is received, it
cancels the request for the received block to all the peers in its
peer set that have the corresponding pending request. As a peer has a
small buffer of pending requests, all blocks are effectively requested
close to the end of the download. Therefore, the \textit{end game
  mode} is used at the very end of the download, thus it has little
impact on the overall performance.

\subsubsection{Choke Algorithm}
\label{sec:choke-algorithm}
The choke algorithm was introduced to guarantee a reasonable level of
upload and download reciprocation. As a consequence, free riders,
i.e., peers that never upload, should be penalized. 
For the sake of clarity, we describe without loss of generality the
choke algorithm from the point of view of the local peer. 
In this section, \textit{interested} always means interested in
the local peer, and \textit{choked} always means choked by the local peer.

The choke algorithm differs in leecher and seed states. 
We describe first the choke algorithm in leecher state. At most 4
remote peers can be unchoked and interested at the same time. Peers
are unchoked using the following policy.
\begin{enumerate}
\item \label{item1LS} 
Every 10 seconds, the interested remote peers
  are ordered according to their download rate to the local peer and
  the 3 fastest peers are unchoked.
\item \label{item2LS} 
 Every 30 seconds, one additional interested remote peer
  is unchoked at random. We call this random unchoke the optimistic
  unchoke.
\end{enumerate}

In the following, we call the three peers unchoked in
step~\ref{item1LS} the regular unchoked (RU) peers, and the peer
unchoked in step~\ref{item2LS} the optimistic unchoked (OU) peer.  The
optimistic unchoke peer selection has two purposes. It allows to
evaluate the download capacity of new peers in the peer set, and it
allows to bootstrap new peers that do not have any piece to share by
giving them their first piece.

We describe now the choke algorithm in seed state.
In previous versions of the \bt protocol, the choke algorithm was the
same in leecher state and in seed state except that in seed state the
ordering performed in step \ref{item1LS} was based on upload rates
from the local peer. With this algorithm, peers with a high download
rate are favored independently of their contribution to the torrent.

Starting with version 4.0.0, the \textit{mainline} client
\cite{btsite} introduced an entirely new algorithm in seed state.  We
are not aware of any documentation on this new algorithm, nor of any
implementation of it apart from the \textit{mainline} client.

We describe this new algorithm in seed state in the following. At most
4 remote peers can be unchoked and interested at the same time. Peers
are unchoked using the following policy.
\begin{enumerate}
\item \label{item1SS} 
 Every 10 seconds, the unchoked and interested
  remote peers are ordered according to the time they were last
  unchoked, most recently unchoked peers first.
\item \label{item2SS} 
 For two consecutive periods of 10 seconds, the 3
  first peers are kept unchoked and an additional
  4\textsuperscript{th} peer that is choked and interested is selected
  at random and unchoked.
\item \label{item3SS} 
 For the third period of 10 seconds, the 4 first
  peers are kept unchoked. 
\end{enumerate}

In the following, we call the three or four peers that are kept unchoked
according to the time they were last unchoked the seed kept unchoked
(SKU) peers, and the unchoked peer selected at random the seed random
unchoked (SRU) peer. With this new algorithm, peers are no longer
unchoked according to their upload rate from the local peer, but
according to the time of their last unchoke. As a consequence, the peers in the
active peer set are changed regularly, each new SRU peer taking an
unchoke slot off the oldest SKU peer.

We show in section~\ref{sec:fairness-issue} why the new
choke algorithm in seed state is fundamental to the fairness of the
choke algorithm.

\section{Experimental Methodology}
\label{sec:exper-meth}
In order to evaluate experimentally the rarest first and
choke algorithms on real torrents, we have instrumented a \bt client
and connected this client to live torrents with different
characteristics. The experiments were performed one at a time in
order to avoid a possible bias due to overlapping experiments. 
We have instrumented a single client and we make no assumption on the
other clients connected to the same torrent. As we only considered
real torrents, we captured a large variety of client configuration,
connectivity, and behavior. In the following, we give details on how
we conducted the experiments.

\subsection{Choice of the Monitored  \bt Client}
\label{sec:choice-bt-client}
Several \bt clients are available. The first \bt client has been
developed by Bram Cohen, the inventor of the protocol. This client is
open source and is called {\itshape mainline} \cite{btsite}. As there is no well
maintained and official specification of the \bt protocol, the
\textit{mainline} client is considered as reference for the \bt
protocol. It should be noted that, up to now, each improvement of Bram
Cohen to the \bt protocol has been replicated to the most popular other
clients.

The other clients differ from the \textit{mainline} client by a more
sophisticated interface with a nice look and feel, realtime
statistics, many configuration options, experimental extensions to the
protocol, etc.

Since our goal is to evaluate the basic \bt protocol, we have
decided to restrict ourselves to the \textit{mainline} client. This
client is very popular as it is the second most downloaded \bt
client at SourceForge with more than 52 million downloads. We
instrumented the version 4.0.2 of the \textit{mainline} client released at
the end of May 2005\footnote{The latest stable branch of development is
  4.20.x. In this branch, there is no new functionality to the core
  protocol, but a new tracker-less functionality and some improvements
  to the client. As the evaluation of the tracker functionality was
  outside the scope of this study we focused on version 4.0.2.}. This
version of the instrumented mainline client implements the new choke
algorithm in seed state (see section~\ref{sec:choke-algorithm}).

\subsection{Choice of the Torrents}
\label{sec:choice-torrent-method}
The aim of this work is to understand how the rarest first and choke
algorithms behave on real torrents. It is not intended to provide an
exhaustive study on the characteristics of today's torrents. For this
reason, we have selected torrents based
on: their proportion of seeds to leechers, the absolute number of
seeds and leechers, and the content size. The torrents monitored in
this study were found on popular
sites\footnote{www.legaltorrents.com, bt.etree.org, fedora.redhat.com, www.mininova.org,
  isohunt.com.}.  We considered copyrighted and free contents,
which are TV shows, movies, cartoons, music albums, live concert
recordings, and softwares. Each experiment lasted for 8 hours in order
to make sure that each client became a seed and to have a
representative trace in seed state. We performed all the experiments
between June 2005 and May 2006. 

We give the characteristic of each torrent in
Table~\ref{table_torrent_charac}. The number of seeds and leechers is
given at the beginning of the experiment. Therefore, these numbers
can be very different at the end of the experiment. 
We see that there is a large variety of torrents: torrents with few
seeds and few leechers, torrents with few seeds and a large number of
leechers, torrents with a large number of seeds and few leechers, and
torrents with a large number of seeds and leechers. 
We discuss in section~\ref{sec:lim-torrent-set} the limitations in the
choice of the torrents considered.

\begin{table}
  \caption{Torrent characteristics.\textit{ \textnormal{\textbf{Column 1
          (ID)}: torrent ID, \textbf{column 2 (\# of S)}:
        number of seeds at the beginning of the experiment, \textbf{column 3
          (\# of L)}: number of leechers  at the beginning of the experiment,
        \textbf{column 4 (Ratio $\frac{S}{L}$)}: ratio (number
        of seeds)/(number of leechers), \textbf{column 5 (Max. PS): }
        maximum peer set size in 
        leecher state, \textbf{column 6 (Size)}: size of the content in MB.}}}
\label{table_torrent_charac}
\begin{center}
\begin{small}
\begin{tabular}{|c|c|c|c|c|c|}
\hline
\textbf{ID} & \textbf{\# of S} & \textbf{\# of L} &
\textbf{Ratio $\frac{S}{L}$}& 
\textbf{Max. PS} &
\textbf{Size}\\
\hline
1 & 0 & 66 & 0 & 60 &  700\\
\hline
2 & 1 & 2 &0.5 & 3 & 580\\ 
\hline
3 &  1 & 29 &0.034& 34 & 350\\ 
\hline
4 & 1 & 40 &0.025 &  75 & 800\\ 
\hline
5 & 1 & 50  &0.02 &  60 & 1419 \\
\hline
6 &  1 & 130 &0.0078 &  80& 820\\ 
\hline
7 &  1 & 713 &0.0014 &  80 & 700\\ 
\hline
8 &  1 & 861 &0.0012 & 80 &  3000\\ 
\hline
9 &  1 & 1055 &0.00095 & 80 &  2000\\ 
\hline
10 &  1 & 1207 &0.00083 &  80 & 348\\ 
\hline
11 &  1 & 1411 &0.00071 &  80 & 710\\ 
\hline
12 & 3 & 612  &0.0049 &  80 & 1413 \\
\hline
13 & 9 & 30 &0.3 & 35 & 350\\ 
\hline
14 & 20 & 126  &0.16 & 80 &  184 \\
\hline
15 &  30 & 230 &0.13 & 80 &  820\\
\hline
16 & 50 & 18 &2.8 & 40 & 600\\
\hline
17 & 102 & 342 &0.3 & 80 & 200\\ 
\hline
18 & 115 & 19 &6 & 55 & 430\\ 
\hline
19 & 160 & 5 &32 & 17 &  6\\
\hline
20 & 177 & 4657 & 0.038 & 80 &  2000\\
\hline
21 &  462 & 180 &2.6 & 80 &  2600\\ 
\hline
22 & 514 & 1703  &0.3 & 80 & 349 \\
\hline
23 & 1197 & 4151 &0.29 & 80 & 349 \\
\hline
24 & 3697 & 7341 &0.5 & 80 & 349 \\
\hline
25 & 11641 & 5418  &2.1 &  80 & 350 \\
\hline
26 &  12612 & 7052 &1.8 &  80 & 140\\
\hline
\end{tabular}
\end{small}
\end{center}
\end{table}

\subsection{Experimental Setup}
\label{sec:exp-setup-method}
We performed a complete instrumentation of the \textit{mainline}
client.
The instrumentation consists of: a log
of each \bt message sent or received with the detailed content of the
message, a log of each
state change in the choke algorithm, a log of the rate estimation
used by the choke algorithm, and a log of important events (end game
mode, seed state). 

As monitored client, we use the \textit{mainline} client with all the default
parameters for all our experimentations. It is outside of the scope of
this study to evaluate the impact of each \bt parameter.
The main default parameters for the monitored client are: the maximum
upload rate (default to 20 kB/s), the minimum number of peers in the
peer set before requesting more peers to the tracker (default to 20),
the maximum number of connections the local peer can initiate (default
to 40), the maximum number of peers in the peer set (default to 80),
the number of peers in the active peer set including the optimistic
unchoke (default to 4), the block size (default to $2^{14}$ Bytes),
the number of pieces downloaded before switching from random to rarest
first piece selection (default to 4).

We did all our experimentations on a machine connected to a high speed
backbone. However, the upload capacity is limited by default by the
client to 20 kB/s. There is no limit to the download capacity. We
obtained effective maximum download speed ranging from 20 kB/s up to
1500 kB/s depending on the experiments. We ran between 1 and 3
experiments on the 26
different torrents given in Table~\ref{table_torrent_charac} and
performed a detailed analysis of each of these traces. The results
given in this paper are for a single run for each torrent. 
Multiple runs on some torrents were used in a calibration phase as
explained in section~\ref{sec:sing-client-intrum}.

Finally, whereas we have control over the monitored \textit{mainline} client,
we do not control any other client in a torrent. In particular, all
peers in the peer set of the local peer are real live peers.

\subsection{Peer Identification}
\label{sec:client-iden-method}
In our experiments, we uniquely identify a peer by its IP address and
peer ID. The peer ID, which is 20 bytes, is a string composed of the
client ID and a randomly generated string. This random string is
regenerated each time the client is restarted. The client ID is a
string composed of the client name and version number, e.g., M4-0-2
for the \textit{mainline} client in version 4.0.2. We are aware of
around 20 different \bt clients, each client existing in several
different versions.  When in a given experiment, we see several peer
IDs corresponding to the same IP address\footnote{Between 0\% and 26\%
  of the IP addresses, depending on the experiments, are associated in
  our traces to more than one peer ID. The mean is around 9\%.}, we
compare the client ID of the different peer IDs. In the case the
client ID is the same for all the peer IDs on a same IP address, we
deem that this is the same peer. We cannot rely on the peer ID
comparison, as the random string is regenerated each time a client
crashes or restarts.  The pair (IP, client ID) does not guarantee that
each peer can be uniquely identified, because several peers beyond a
NAT can use the same client in the same version. However, considering
the large number of client IDs, it is common in our experiments to
observe 15 different client IDs, the probability to have several different
clients beyond a NAT with the same client ID is reasonably low for our
purposes.  Moreover, unlike what was reported by Bhagwan et al. \cite{bhagwan03}
for the Overnet file sharing network, we did not see any problem of
peer identification due to NATs. In fact, \bt has an option, activated
by default, to prevent accepting multiple concurrent incoming
connections from the same IP address. The idea is to prevent peers to
increase their share of the torrent, by opening multiple clients from
the same machine. Therefore, even if we found in our traces different
peers with the same IP address at different moments in time, two
different peers with the same IP address cannot be connected to the
local peer during overlapping periods.

\subsection{Limitations and Interpretation of the Results}
\label{sec:method-lim}
In this section we discuss the two main limitations of this work, namely the
single client instrumentation and the limited set of monitored
torrents.  We also discuss why, despite these limitations, we believe
our conclusions hold for a broader range of scenarios than the ones
presented. 

\subsubsection{Single Client Instrumentation}
\label{sec:sing-client-intrum}
We have chosen for this study to focus on the behavior of a single
client in a real torrent. Whereas it may be argued that a larger
number of instrumented peers would have given a better understanding
of the torrents, we made the decision to be as unobtrusive as
possible. Increasing the number of instrumented clients would have
required to either control those clients ourselves, or to ask some
peers to use our instrumented client. In both cases, the choice of the
instrumented peer set would have been biased, and the behavior of the
torrent impacted. Instead, our decision was to understand how
a new peer (our instrumented peer) joining a real torrent behaves.

Moreover, monitoring a single client does not adversely impact the
generality of our findings for the following reasons. First, a torrent
is a random graph of interconnected peers. For this reason, with a
large peer set of 80, each peer should have a view of the torrent as
representative as any other peer. Even if each peer will see
variations due to the random choice of the population in its peer set,
the big picture will remain the same. Second, in order to make sure
that there is no unforeseen bias due to the single client
instrumentation, we have monitored several torrents with three
different peers, each peer with a different IP address. These
experiments were performed during a calibration phase, and are not
presented here due to space limitation.  Whereas the download speed of
the peers may significantly vary, e.g., due to very fast seeds that
may of may not be present in the peer set of a monitored client, we
did not observe any other significant difference among the clients
that may challenge the generality of our findings.

\subsubsection{Limited Torrent Set}
\label{sec:lim-torrent-set}
We have considered for this study 26 different torrents. Whereas it is
a large number of torrents, it is not large enough to be exhaustive or
to be representative of all the torrents that can be found in the
Internet. However, our intent is to evaluate the behavior of the
rarest first and choke algorithm in a variety of situations. The
choice of the torrents considered in this study was targeted to
provide a challenging environment to the rarest first and choke
algorithms. For instance, torrents with no seed (torrent 1) or with
only one seed and a large number of leechers (e.g., torrent 7--11)
were specifically chosen to evaluate how the rarest first algorithm
behaves in the context of pieces scarcity. Torrents with a large number
of peers were selected to evaluate how the choke algorithm behaves when
the torrent is large enough to favor free riders.

We have around half of the presented torrents with no or few seeds, as
this is a challenging situation for a \p2p protocol.  However, it can
be argued that the largest presented torrent with a single seed has a
small number of leechers (1441 leechers at the beginning of the
experiment for torrent 11). Indeed, the target of a \p2p protocol is
to distribute content to millions of peers. But, a \p2p protocol
capitalizes on the bandwidth of each peer. Thus, it is not possible to
scale to millions of peers without a significant proportion of seeds.
If we take the same proportion of seeds and leechers as the one of torrent 11,
only 710 seeds are enough to scale to one million of peers. Also, a
torrent with a ratio $\frac{\mbox{number of seeds}}{\mbox{number of
    leechers}}$ lower than $10^{-3}$ is enough to stress a piece
selection strategy based on a local view of only 80 peers.

Finally, in such an experimental study it is not possible to reproduce
an experiment, and thus to gain statistical information because each
experiment depends on the behavior of peers, the number of seeds and
leechers in the torrent, and the subset of peers randomly returned by
the tracker.  However, studying the dynamics of the protocol is as
important as studying its statistical properties. As we considered
torrents with different characteristics and observed a consistent
behavior on these torrents, we believe our findings to be
representative of the rarest first and choke algorithms behavior.

\section{Experimental Results}
\label{sec:experimentation-results}
We present in this section the results of our experiments. In a first
part, we discuss the results with a focus on the rarest first
algorithms. Then, in a second part, we discuss the results with
a focus on the choke algorithm.

\subsection{Rarest First Algorithm}
\label{sec:rarest-first-algor}
The aim of a piece selection strategy is to guarantee that each peer
is always interested in any other peer. The rational is that each time
the peer selection strategy unchokes a peer, this peer must be
interested in the unchoking peer. This way, the peer selection
strategy can reach the optimal system capacity (but, designing such an
optimal peer selection strategy is a hard task). Therefore, the piece
selection strategy is fundamental to reach good system capacity.

However, the efficiency of the piece selection strategy cannot be measured in terms of
system capacity, because the system capacity is the result of both the
piece and peer selection strategies. A good way to evaluate the
efficiency of the piece
selection strategy is to measure the \textit{entropy} of the
torrent, i.e., the repartition of pieces among peers.

There is no simple way to directly measure the entropy of a torrent.
For this reason, we characterize the entropy with the \textit{peer
  availability}. We define the peer availability of peer $x$ according
to peer $y$ as the ratio of the time peer $y$ is interested (see
section~\ref{sec:terminology}) in peer $x$ over the time peer $x$ is
in the peer set of peer $y$. If peer $x$ is always available for peer
$y$, then the peer availability is equal to one. In the following, we
characterize the entropy of a torrent with the availability of the
peers in this torrent. For the sake of clarity, we will simply refer
to the notion of entropy.

We say that there is ideal entropy in a torrent when each
leecher\footnote{Only the case of leechers is relevant for the entropy
  characterization, as seeds are always interesting for leechers and
  never interested in leechers.} is always interested in any other
leecher. We do not claim that ideal entropy can be always achieved,
but it should be the objective of any efficient piece selection
strategy. 

We evaluated the rarest first algorithm on a representative set of
real torrents. We showed that the rarest first algorithm achieves a
close to ideal entropy, and that its replacement by more
complex solutions cannot be justified.  Then, we evaluated the
dynamics of the rarest first algorithm to understand the reasons for
this good entropy. Finally, we focused on a specific problem called
the last pieces problem, which is presented \cite{gkantsiInf05,
  kosticUsenix05} as a major weakness of the rarest first strategy. We
showed that the last pieces problem is overestimated. In contrast, we
identified a first blocks problem, which is a major area of
improvement for \btns.

\subsubsection{Entropy Characterization}
\label{sec:entropy-charac}
\begin{figure}
\centering
\includegraphics[width=3.1in]{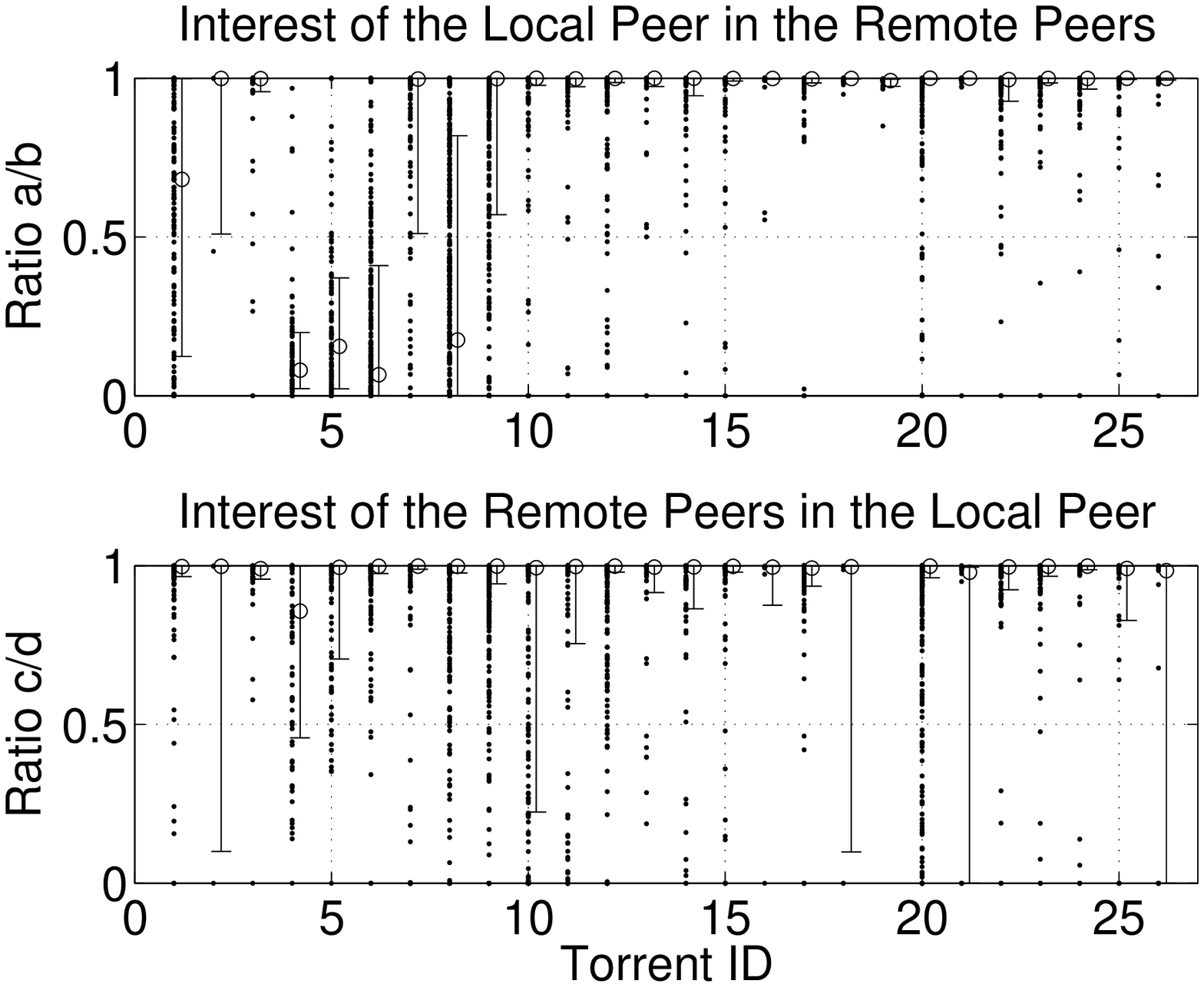}
\caption{\textmd{\textsl{Entropy characterization. \textbf{Top graph:} For
  each remote leecher peer for a given torrent, a dot represents the
  ratio $\frac{a}{b}$ where $a$ is the time the local peer in leecher
  state is interested in this remote peer and $b$ is the time this
  remote peer spent in the peer set when the local peer is in leecher
  state.  \textbf{Bottom graph:} For each remote leecher
  peer for a given torrent, a dot represents the ratio $\frac{c}{d}$
  where $c$ is the time this remote peer is interested in the local
  peer in leecher state and $d$ is the time this remote peer spent in
  the peer set when the local peer is in leecher state.
  \textbf{For both graphs:} Each vertical solid lines
  represent the 20\textsuperscript{th} percentile (bottom of the
  line), the median (identified with a circle), and the
  80\textsuperscript{th} percentile (top of the line) of the ratios
  for a given torrent. }}}
\label{fig:entropy-charac}
\end{figure}

The major finding of this section is that the rarest first algorithm
achieves a close to ideal entropy for real torrents. We remind that
ideal entropy is achieved when each leecher is always interested in
any other leecher. As we do not have global knowledge of the torrent,
we characterize the entropy from the point of view of the local peer
with two ratios. For each remote peer we compute:
\begin{itemize}
\item the ratio $\frac{a}{b}$ where $a$ is the time the local peer in leecher
  state is interested in this remote peer and $b$ is the time this
  remote peer spent in the peer set when the local peer is in leecher
  state;
\item the ratio $\frac{c}{d}$
  where $c$ is the time this remote peer is interested in the local
  peer in leecher state and $d$ is the time this remote peer spent in
  the peer set when the local peer is in leecher state.
\end{itemize}
In the case of ideal entropy the above ratios should be one. 
Fig.~\ref{fig:entropy-charac} gives a characterization of the entropy
for the torrents considered in this study.  

For most of our
torrents, we see in Fig.~\ref{fig:entropy-charac} that the ratios
are close to 1, thus a close to ideal entropy.
For the top graph, 70\% of the torrents have the
20\textsuperscript{th} percentile close to one, and 80\% have the
median close to one. For the bottom
graph, 70\% of the torrents have a 20\textsuperscript{th} percentile
close to one, and 90\% of the torrents have the median close to one.  We
discuss below the case of the torrents with low entropy.

First, we discuss why the local peer is often not interested in the
remote peers for torrents 1, 2, 4, 5, 6, 7, 8, and 9 (see
Fig.~\ref{fig:entropy-charac}, top graph). These torrents have low
entropy because they are in a startup phase. This means that the
initial seed has not yet served all the pieces of the content. We
remind that the pieces only present on the initial seed are the
\textit{rare pieces}, and that the pieces already served at least once
by the initial seed are the \textit{available pieces} (see
section~\ref{sec:terminology}).
The reason for the low observed entropy is that during a torrent
startup, available pieces are replicated with an exponential capacity
of service \cite{yang04}, but rare pieces are served by the initial
seed at a constant rate. Thus, available pieces are replicated faster
than rare pieces. This leads to two problems.  First, the probability
of having peers in a peer set with the same subset of pieces is higher
during the torrent startup than when there is no rare piece in the
torrent.  Second, when there is no rare piece, a peer with all the
available pieces becomes a seed. But, when there are rare pieces, a peer
with all the available pieces remains a leecher because it does not
have the rare pieces. However, these leechers cannot be interested in
any other peer as they have all the available pieces at this point of
time, but they stay in the peer set of the local peer. Thus a low
ratio for these leechers in Fig.~\ref{fig:entropy-charac}.  In conclusion, the low
entropy we observed is not due to a deficiency of the rarest first
algorithm, but to the startup phase of the torrent whose duration
depends only on the upload capacity of the initial seed. We discuss further
this point in section~\ref{sec:transient-state}.

Now, we discuss why the remote peers are often not interested in the
local peer for torrents 2, 4, 10, 18, 19, 21, and 26 (see
Fig.~\ref{fig:entropy-charac}, bottom graph). No dot is displayed for
torrent 19 because due to the small number of leechers in this torrent,
the local peer in leecher state had no leecher in its peer
set. Five torrents have a 20\textsuperscript{th} percentile close to
0. The percentile for four of these torrents is computed on a small number of
ratios: 3, 8, 12, and 15 for torrents 2, 18, 21 and 26 respectively.
Therefore, the 20\textsuperscript{th} percentile is not representative
as it is not computed on a set large enough. Additionally, the reason
for the low 20\textsuperscript{th} percentile is peers with a ratio of
0. We identified two reasons for a ratio of 0. First, some peers
join the peer set with almost all pieces. They are therefore unlikely
to be interested in the local peer. Second, some peers with no or few
pieces never sent an interested message to the local peer. This can be
explained by a client behavior changed with a plugin or an option
activation. The super seeding option \cite{btwikispec} available in
several \bt clients has this effect. In conclusion, the low entropy
of some peers is either a measurement artifact due to modified or
misbehaving clients, or the result of the inability of the rarest
first algorithm to reach ideal entropy in some extreme cases.

We have seen that peers that join the torrent with almost all pieces
may not be interested in the local peer. In this scenario, the rarest first
algorithm does not guarantee ideal entropy.  However, we argue that 
this case does not justify the replacement of the rarest first 
algorithm for two reasons. First, this case appears rarely
and does not significantly impact the overall entropy of the
torrent. Second, the peers with low entropy are peers that join the
peer set with only a few missing pieces. In the case of torrent startup,
it is not clear whether a
solution based, for instance, on source or network coding would have
proposed interesting pieces to such peers. Indeed, when content is
split into $k$ pieces, there is no solution based on coding that can
reconstruct the content in less than $k$ pieces. For this reason, when
the initial seed has not yet sent at least one copy of each piece, there is
no way to reconstruct the content, so no way to have interesting
pieces for all the peers. 

An important question is how rarest first compares with network coding
in the presented scenarios. As there is no client based on network
coding that is as popular as \btns, it is not possible to evaluate
both solutions on the same torrents. However, based on the theoretical
network coding results, we discuss the respective merits of rarest
first and network coding in section~\ref{sec:disc-network-cod}.

For the computation of the ratios on Fig.~\ref{fig:entropy-charac}, we
did not consider peers that spent less than 10 seconds in the peer
set. Our motivation was to evaluate the entropy of pieces in a
torrent. However, due to several misbehaving clients, there is a
permanent noise created by peers that join and leave the peer set
frequently. Such peers stay typically less than a few seconds in the
peer set, and they do not take part in any active upload or download.
Therefore, these misbehaving peers adversely bias our entropy
characterization. Filtering all peers that stay less than 10 seconds
remove the bias.

In summary, we have seen that the rarest first algorithm enforces a
close to ideal entropy for the presented torrents. We have
identified torrents with low entropy and shown that the rarest
first algorithm is not responsible for this low entropy. We have also
identified rare cases where the rarest first algorithm does not
perform optimally, but we have explained that these cases do not
justify a replacement with a more complex solution.  In the following,
we evaluate how the rarest first piece selection strategy achieves 
high entropy.

\subsubsection{Rarest First Algorithm Dynamics}
\label{rarest-first-alg-dyn}
We classify a torrent in two states: the transient state and the steady
state\footnote{Our definition of transient and steady state differs
  from the one given by Yang et al. \cite{yang04}.}.  In transient
state, there is only one seed in the torrent. In particular, there are
some pieces that are rare, i.e., present only at the seed. This state
corresponds to the beginning of the torrent, when the initial seed has not
yet uploaded all the pieces of the content. All torrents with 
low entropy (Fig.~\ref{fig:entropy-charac}, top graph) are in 
transient state. A good piece replication algorithm should minimize
the time spent in the transient state because low entropy may
adversely impact the service capacity of a torrent by biasing the peer
selection strategy.  In steady state, there is no rare piece, and the
piece replication strategy should prevent the torrent to enter again
a transient state. All torrents with high entropy are in steady
state.

In the following, we evaluate how the rarest first algorithm performs
in transient and steady state. We show that the low entropy of
torrents experienced in transient state is due to the limited upload
capacity of the initial seed, and that the rarest first algorithm
minimizes the time spent in this state. We also show that the rarest
first algorithm is efficient at keeping a torrent in steady state,
thus guaranteeing a high entropy.

\paragraph{Transient State}
\label{sec:transient-state}
In order to understand the dynamics of the rarest first algorithm in
transient state, we focus on torrent 8. This torrent consisted of 1
seed and 861 leechers at the beginning of the experiment. The file
distributed in this torrent is split in 863 pieces. We run this
experiment during 58991 seconds, but in the following we only discuss
the results for the first 29959 seconds when the local peer is in
leecher state.

\begin{figure}
\centering
\includegraphics[width=2.7in]{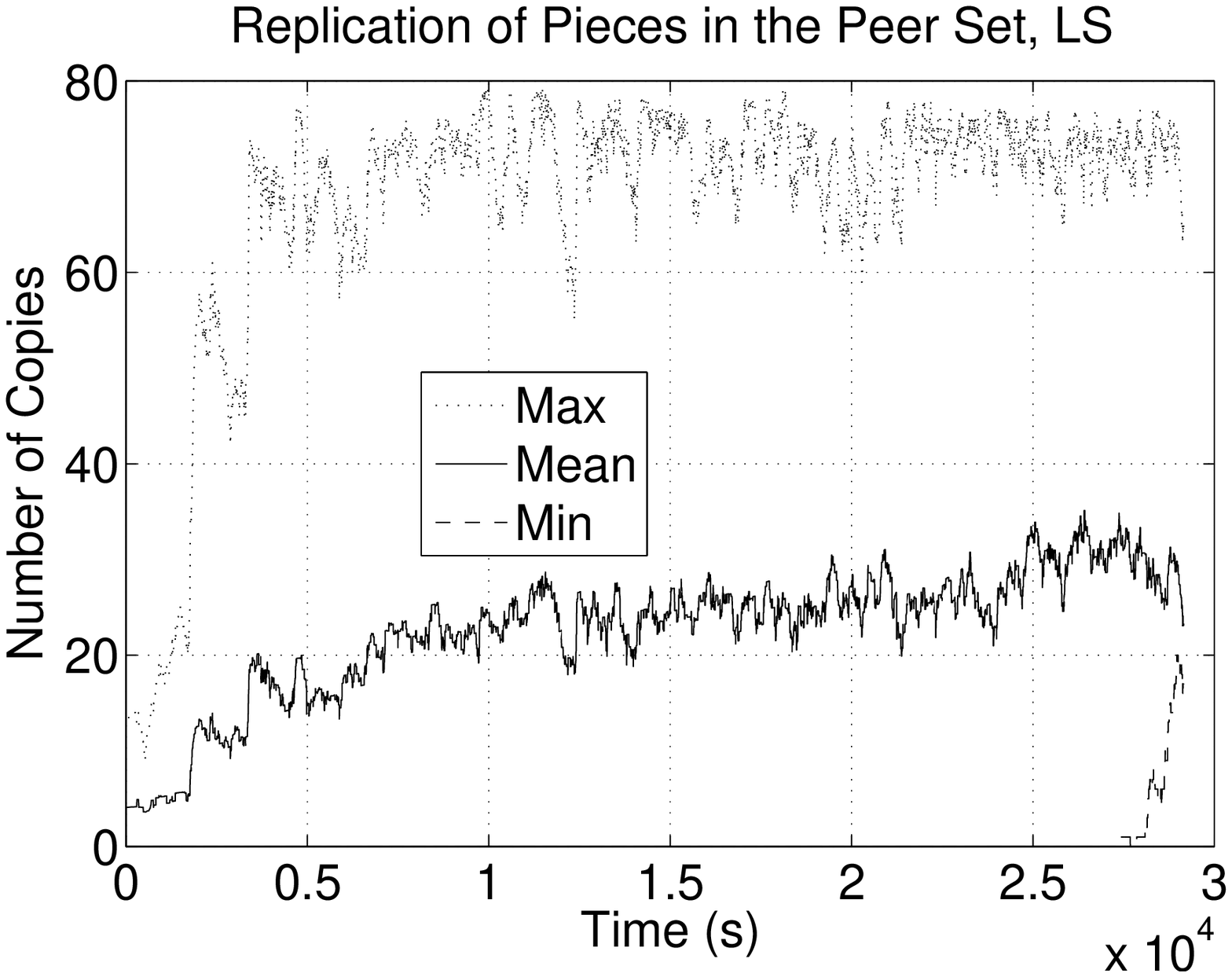}
\caption{\textmd{\textsl{Evolution of the number of copies of pieces in the peer set
  with time for torrent 8 in leecher state. \textbf{Legend:} The dotted line represents the number of
  copies of the most replicated piece in the peer set at each instant.
  The solid line represents the mean number of copies over all the
  pieces in the peer set at each instant. The dashed line represents
  the number of copies of the least replicated piece in the peer set
  at each instant.}}}
\label{fig:num-copies-noseed}
\end{figure}

Torrent 8 is in transient state for most of the experiment. As we
don't have global knowledge of the torrent, we do not have a direct
observation of the transient state. However, there are several
evidences of this state. Indeed, Fig.~\ref{fig:num-copies-noseed} shows
that there are missing pieces during the experiment in the local peer
set, as the minimum curve (dashed line) is at zero.  Moreover, we
probed the tracker to get statistics on the number of seeds and
leechers during this experiment. We found that this torrent had only
one seed for the duration of the experiment.

\begin{figure}
\centering
\includegraphics[width=2.7in]{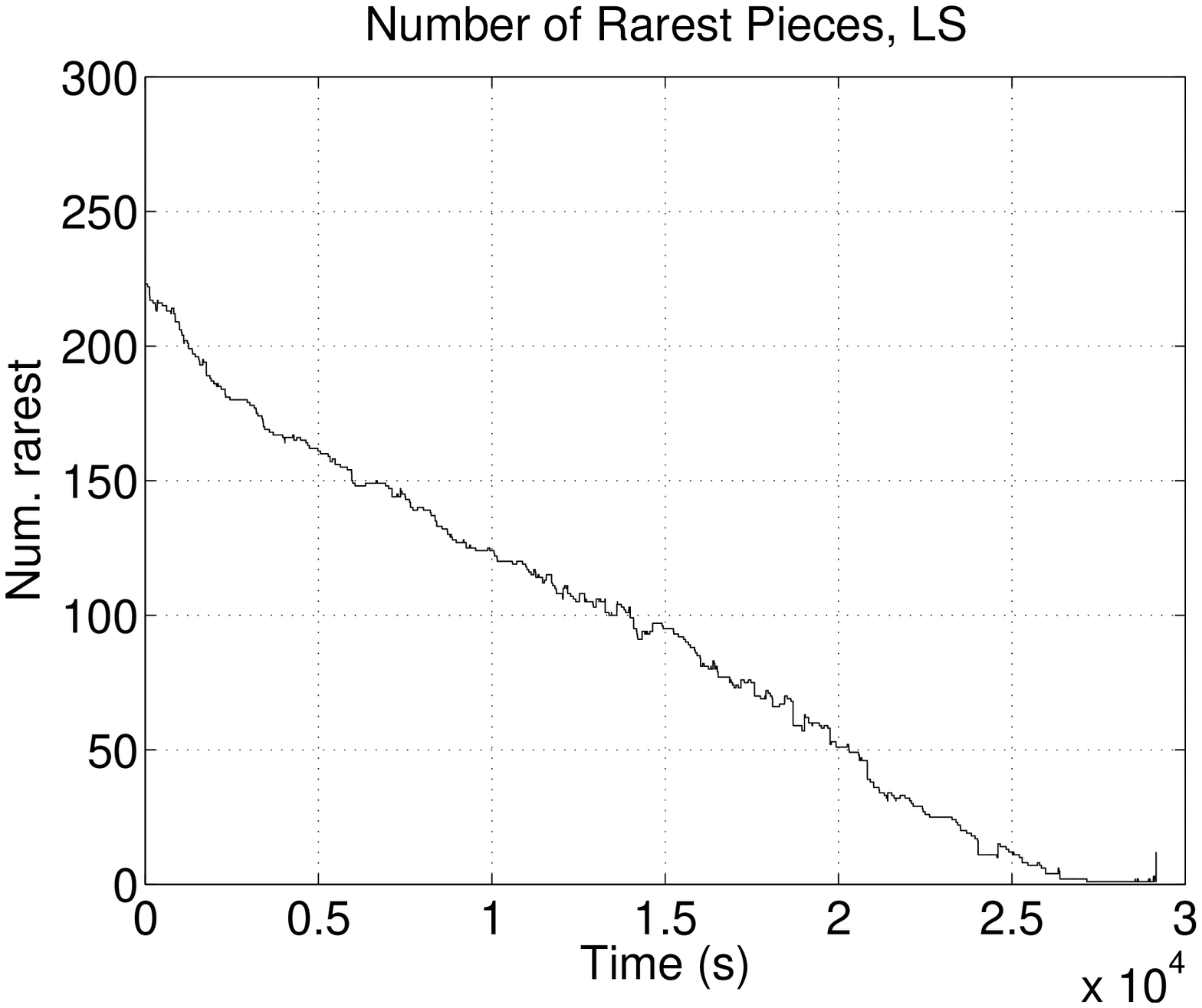}
\caption{\textmd{\textsl{Evolution of the number of rarest pieces in the peer set for
torrent 8 in leecher state.  The rarest pieces set is formed by the pieces that are equally the
rarest, i.e., the pieces that have the least number of copies in the peer set.}}}
\label{fig:num-rarest-noseed}
\end{figure}

We see in Fig.~\ref{fig:entropy-charac}, top graph, that torrent 8
has low entropy. This low entropy is due to the limited upload
capacity of the initial seed. Indeed, when a torrent is in transient
state, available pieces are replicated with an exponential capacity of
service \cite{yang04}, but rare pieces are served by the initial seed
at a constant rate.  This is confirmed by
Fig.~\ref{fig:num-rarest-noseed} that shows the number of rarest
pieces, i.e., the set size of the pieces that are equally rarest.
We see that the number of rarest pieces decreases linearly with time.
As the size of each piece in this torrent is 4 MB, a rapid calculation
shows that the rarest pieces are duplicated in the peer set at a
constant rate close to 36 kB/s. We do not have a direct proof that
this rate is the one of the initial seed, because we do not have
global knowledge of the torrent. However, the torrent is in its
startup phase and most of the pieces are only available on the initial
seed. Indeed, Fig.~\ref{fig:num-copies-noseed} shows that there are
missing pieces in the peer set, thus the rarest pieces presented in
Fig.~\ref{fig:num-rarest-noseed} are missing pieces in the peer set.
Therefore, only the initial seed can serve the missing pieces shown in
Fig.~\ref{fig:num-rarest-noseed}.  In conclusion, the upload capacity
of the initial seed is the bottleneck for the replication of the rare
pieces, and the time spent in transient state only depends on the
upload capacity of the initial seed.

The rarest first algorithm attempts to minimize the time spent in
transient state and replicates fast available pieces. Indeed, leechers
download first the rare pieces. As the rare pieces are only present on
the initial seed, the upload capacity of the initial seed will be
fully utilized and no or few duplicate rare pieces will be served by
the initial seed.
Once served by the initial seed, a rare piece becomes available and is
served in the torrent with an increasing capacity of service. As rare
pieces are served at a constant rate, most of the capacity of service
of the torrent is used to replicate the available pieces on leechers.
Indeed, Fig.~\ref{fig:num-copies-noseed} shows that once a piece is
served by the initial seed, the rarest first algorithm will start to
replicate it fast as shown by the continuous increase in the mean
number of copies over all the peers, and by the number of copies of
the most replicated piece (dotted line) that is always close to the
maximum peer set size of 80.

In summary, the low entropy observed for some torrents is due to the
transient phase. The duration of this phase cannot be shorter than the
time for the initial seed to send one copy of each piece, which is
constrained by the upload capacity of the initial seed. Thus, the time
spent in this phase cannot be shorten further by the piece replication
strategy.  The rarest first algorithm minimizes the time spent in
transient state. Once a piece is served by the initial seed, the
rarest first algorithm replicates it fast. Therefore, a replacement of
the rarest first algorithm by another algorithm cannot be justified
based on the real torrents we have monitored in transient state.

\begin{figure}
\centering
\includegraphics[width=2.7in]{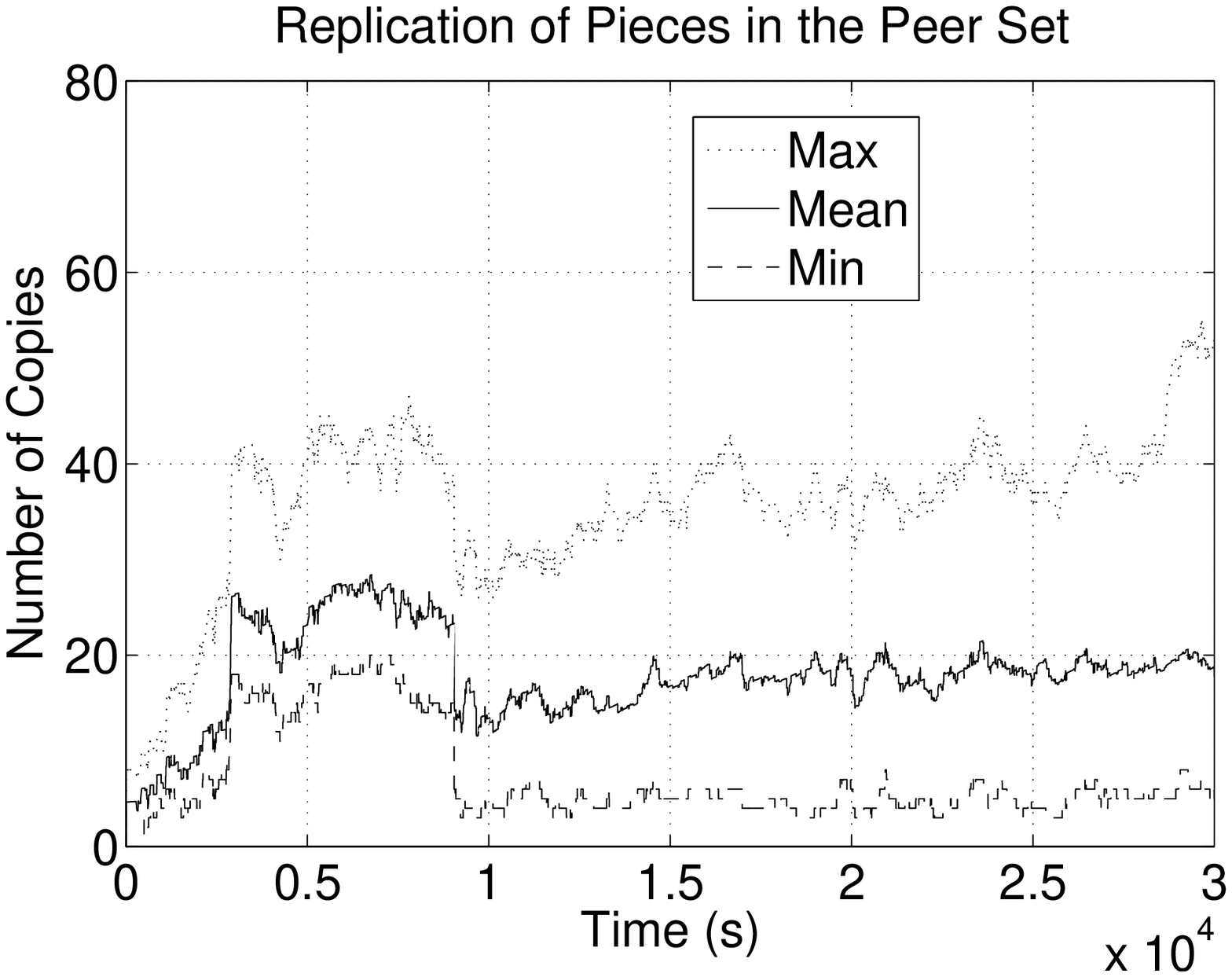}
\caption{\textmd{\textsl{Evolution of the number of copies of pieces in the peer set
  with time for torrent 7. \textbf{Legend:} The dotted line represents the number of
  copies of the most replicated piece in the peer set at each instant.
  The solid line represents the mean number of copies over all the
  pieces in the peer set at each instant. The dashed line represents
  the number of copies of the least replicated piece in the peer set
  at each instant. }}}
\label{fig:num-copies}
\end{figure}

\paragraph{Steady State}
\label{sec:steady-state}
In order to understand the dynamics of the rarest first algorithm in
steady state, we focus on torrent 7.  This torrent consisted of 1 seed
and 713 leechers at the beginning of the experiment. We have seen on
Fig.~\ref{fig:entropy-charac} that torrent 7 has a high entropy.
Fig.~\ref{fig:num-copies} shows that the least replicated piece (min
curve) has always more than 1 copy in the peer set. Thus, torrent 7 is
in steady state.

\begin{figure}
\centering
\includegraphics[width=2.7in]{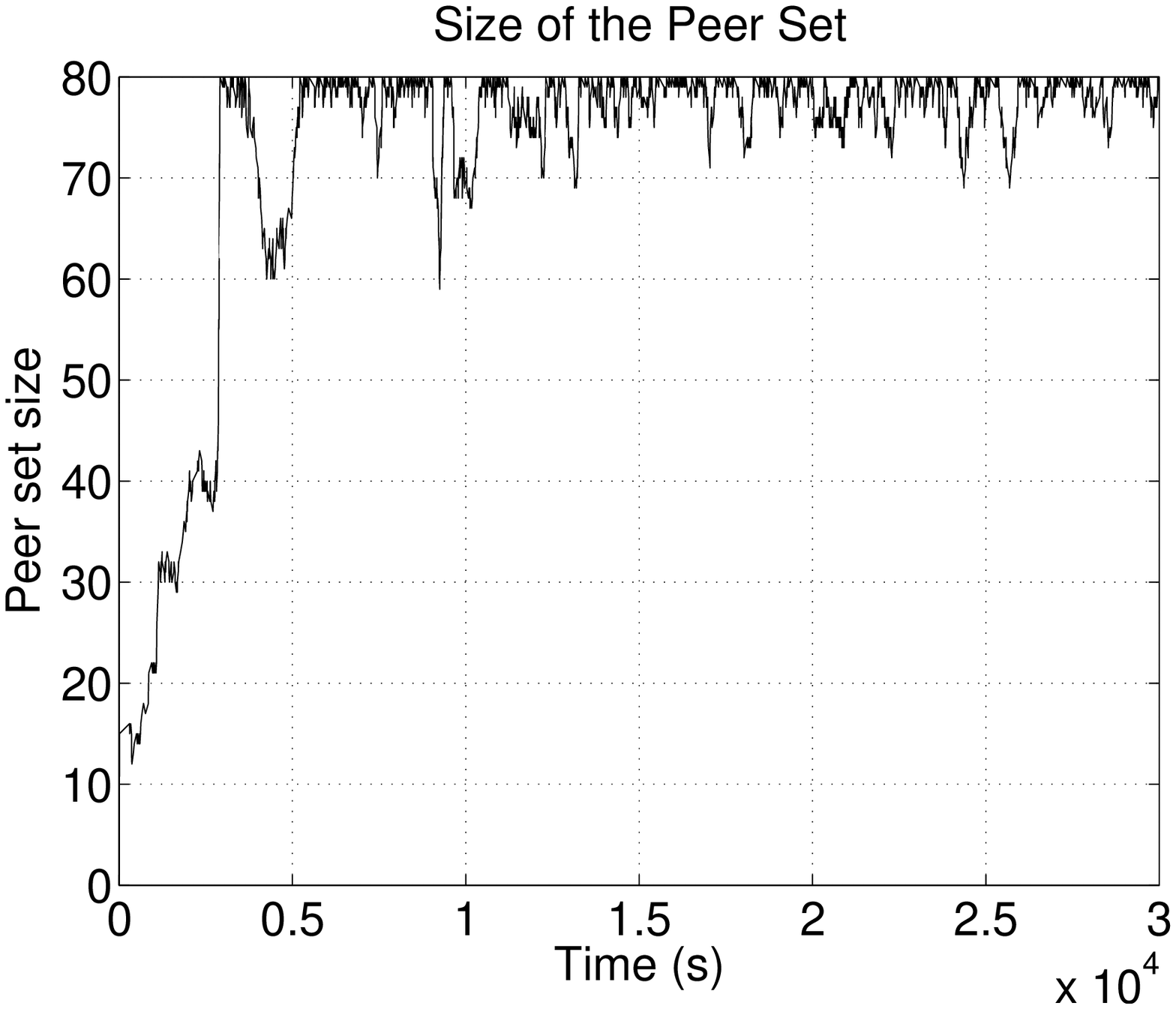}
\caption{\textmd{\textsl{Evolution of the peer set size for torrent 7.}}}
\label{fig:peer-set}
\end{figure}

In the following, we present the dynamics of the rarest first
algorithm in steady state, and explain how this algorithm prevents the
torrent to return in transient state.  Fig.~\ref{fig:num-copies} shows
that the mean number of copies remains well bounded over time by the
number of copies of the most and least replicated pieces. The
variation observed in the number of copies are explained by the
variation of the peer set size, see Fig.~\ref{fig:peer-set}.  The
decrease in the number of copies 9051 seconds after the beginning of
the experiment corresponds to the local peer switching to seed state.
Indeed, when a leecher becomes a seed, it closes its connections to
all the seeds.

\begin{figure}
\centering
\includegraphics[width=2.7in]{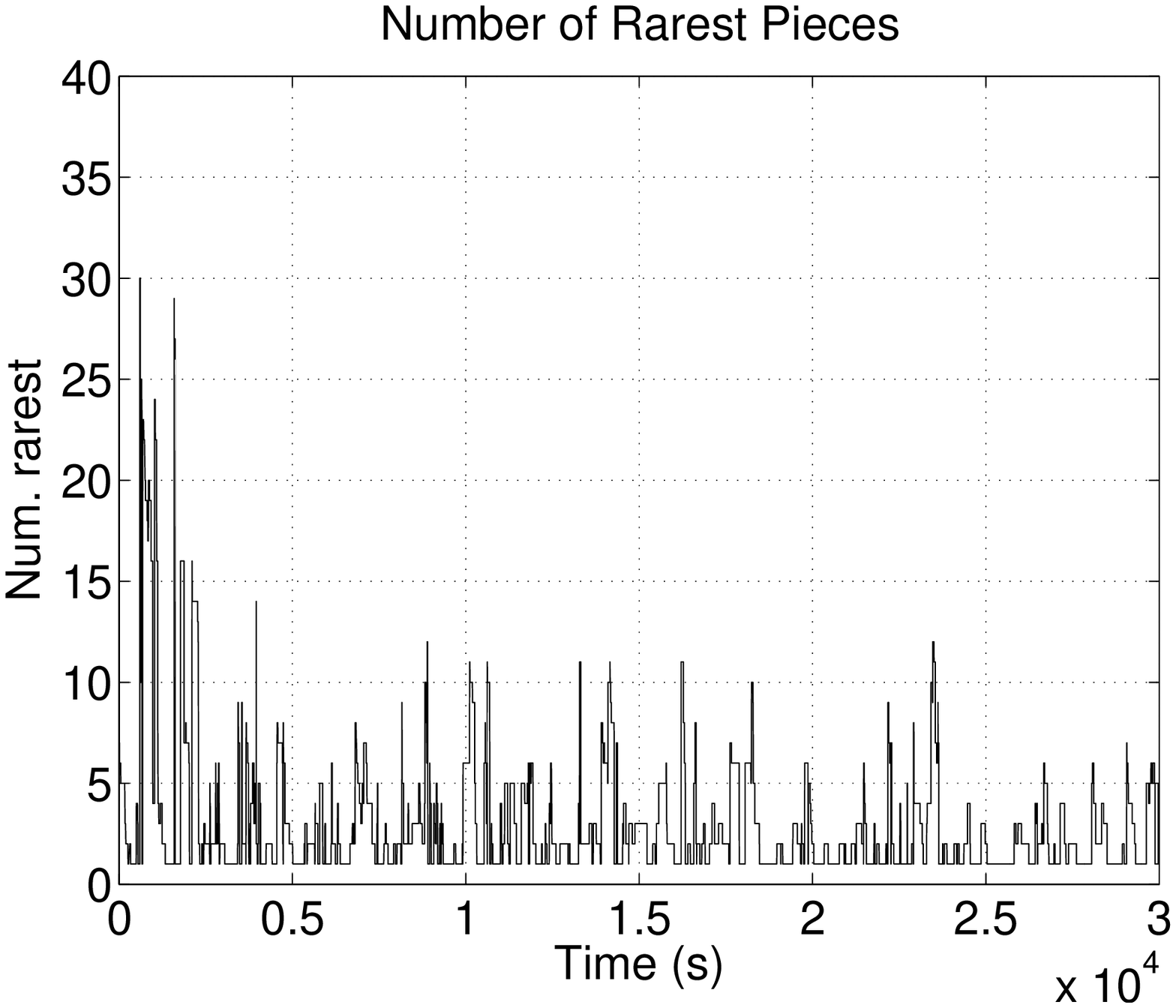}
\caption{\textmd{\textsl{Evolution of the number of rarest pieces in the peer set for
  torrent 7. The rarest pieces set is formed by the pieces that are equally the
rarest, i.e., the pieces that have the least number of copies in the peer set.}}}
\label{fig:num-rarest}
\end{figure}

The rarest first algorithm does a very good job at increasing the
number of copies of the rarest pieces.  Fig.~\ref{fig:num-copies}
shows that the number of copies of the least replicated piece (min
curve) closely follows the mean, but does not significantly get
closer. However, we see in Fig.~\ref{fig:num-rarest} that the number
of rarest pieces, i.e., the set size of the pieces that are equally
rarest, follow a sawtooth behavior. Each peer joining or leaving the
peer set can alter the set of rarest pieces.  But, as soon as a new
set of pieces becomes rarest, the rarest first algorithm quickly
duplicates them as shown by a consistent drop in the number of rarest
pieces in Fig.\ref{fig:num-rarest}.  Finally, we never observed in any
of our torrents a steady state followed by a transient state.

In summary, the rarest first algorithm in steady state ensures a good
replication of the pieces in real torrents. It also replicates fast
the rarest pieces in order to prevent the reappearance of a transient
state.  We conclude that on real torrents in steady state, the rarest
first algorithm is enough to guarantee a high entropy.

\subsubsection{Last Pieces Problem}
\label{sec:last-piece-problem}
We say that there is a last pieces\footnote{This problem is usually
  referenced as the last piece (singular) problem. However, there is
  no reason why this problem affects only a single piece.} problem
when the download speed suffers a significant slow down for the last
pieces. This problem is due to some pieces replicated on few
overloaded peers, i.e., peers that receive more requests than they can serve.
This problem is detected by a peer only at the end of the
content download. Indeed, a peer always seeks for fast peers to
download from. Thus, it is likely that if some pieces are available on
only few overloaded peers, these peers will be chosen only at the end
of the content download when there is no other pieces to download. 

Due to space limitation, we just give our main conclusions. For a
detailed discussion on the last pieces problem, the interested reader
may refer to \cite{legoutBT06}. 

We never observed a last pieces problem on torrents in steady state.
However, we observed this problem on a few torrents in transient
state. We found that this problem is inherent to the transient state
of the torrent, and is not due to the rarest first algorithm.
Moreover, the rarest first algorithm is efficient at mitigating this
problem by replicating fast rare pieces once they become available.

It is important to study the piece interarrival time, because
partially received pieces cannot be retransmitted by a \bt client,
only complete pieces can. However, pieces are split into blocks, which
are the \bt unit of data transfer. For this reason, we have also
evaluated the block interarrival time.  We identified a first blocks
problem. This first blocks problem results in a slow startup of the
torrent, which is an area of improvement for \btns.

In conclusion, the last pieces problem is overstated, but the first
blocks problem is underestimated and a possibility of performance
improvement.

\subsubsection{Discussion on Rarest First and Network Coding}
\label{sec:disc-network-cod}
We have seen that rarest first is an efficient piece selection
strategy on the presented torrents. We have also shown that the
claimed deficiencies of rarest first cannot be identified in our
experiments, or are the results of a misunderstanding of the reason of
piece scarcity for torrents in transient state. 

However, this paper is not a case against solutions based on source or
network coding. Network coding enables a piece selection strategy that
is close to optimal in all cases, which is not the case of rarest
first. Indeed, in specific contexts like small outdegree constraint, or
poor network connectivity between cluster of peers, rarest first will
perform poorly. In this study, we show that on real torrents in the Internet, which
have a large peer set of 80 and do not suffer from connectivity problems,
rarest first performs very well. 

In fact, rarest first is close to a solution based on network coding
in the presented torrents. We consider two cases to make the
comparison: the steady and transient states. In steady state, we have
seen in section~\ref{sec:steady-state} that the entropy of the
presented torrents is close to one with rarest first. An entropy close
to one means that each peer is interested in each other peer in its
peer set most of the time. As this is close to the target of an ideal
piece selection strategy, we see that in steady state, the possibility
of improvement for any piece selection strategy in not significant
compared to rarest first. For this reason, we argue that a replacement
of rarest first cannot be justified in the studied context.  In
transient state, a solution based on network coding will enable the
initial seed to send one entire copy of the content faster than in the
case of rarest first that may suffer from duplicate pieces. The
problem with rarest first is that the number of duplicate pieces will
depends on the peer selection strategy. Indeed, if the initial seed
chooses the same set of peers to upload the initial pieces to and that
these peers are all in the same peer set, then they will have the same
view of the rarest pieces, and they will download from the initial
seed an entire copy of the content without any duplicate pieces. But,
other peer selection policies may increase the ratio of duplicate
pieces before a first copy of the content is sent. There is no such a
problem with network coding. However, simple policies can be
implemented to guarantee that the ratio of duplicate pieces remains
low for the initial seed, e.g., the new choke algorithm in seed state
or the super seeding mode \cite{btwikispec}. In this case, the benefit
of network coding compared to rarest first will not be significant at
the scale of the content download.

Network coding appears as a solution more general than rarest
first, as it works optimally in all cases. However, we argue in
favor of the simplicity of
rarest first. Network coding raises
several implementation issues and is CPU intensive. Rarest first is
simple, easy to implement, and already widely used. We have seen that
in a context of \p2p content replication with a large peer set and 
a good network connectivity, rarest first is a simple and very
efficient solution. That is in this context that we argue that a
replacement of rarest first cannot be justified.

\subsection{Choke Algorithm}
\label{sec:choke-algorithm-1}
The choke algorithm is a peer selection strategy. It should guarantee
fairness and maximize the system capacity. In this section, we focus
on the fairness issue, as the claimed deficiencies of the choke
algorithm are related to its fairness properties. Whereas the
evaluation and optimization of the system capacity is an important
issue, the choke algorithm is indisputably an efficient peer selection
strategy that is used by millions of persons. A detailed evaluation of
the system capacity reached with the choke algorithm is an interesting
area of future research.

\subsubsection{Fairness Issue}
\label{sec:fairness-issue}
Several recent studies
\cite{GuoIMC05,JunSigWork05,Ganesan05,bharambeApr06} challenge the
fairness properties of the choke algorithm because it does not
implement a bit level tit-for-tat, but a coarse approximation based on
short term download estimations. Moreover, it is believed that a fair
peer selection strategy must enforce a byte level reciprocation. For instance,
a peer $A$ refuses to upload data to a peer $B$ if the amount of bytes
uploaded by $A$ to $B$ minus the amount of bytes downloaded from $B$
to $A$ is higher than a given threshold
\cite{JunSigWork05,Ganesan05,bharambeApr06}.  The rationale behind
this notion of fairness is that free riders should be penalized, and
reciprocation should be enforced. We call this notion of fairness,
tit-for-tat fairness.

We argue in the following that tit-for-tat fairness is not appropriate
in the context of \p2p file replication.
A \p2p session consists of seeds, leechers, and free riders, i.e.,
leechers that never upload data. We consider the free
riders as a subset of the leechers. With tit-for-tat fairness, when
there is more capacity of service in the torrent than request for
this capacity, the excess capacity will be lost even if slow leechers
or free riders could benefit from it. Excess capacity is not rare as
it is a fundamental property of \p2p applications. Indeed, there are two
important characteristics of \p2p applications that tit-for-tat
fairness does not take into account.  First, leechers can have an
asymmetrical network connectivity, the upload capacity being lower than
the download capacity. In the case of tit-for-tat fairness, a leecher
will never be able to use its full download capacity even if there is
excess capacity in the \p2p session.  
Second, a seed cannot evaluate the reciprocation of a leecher, because
a seed does not need any piece. As a consequence, there is no way for
a seed to enforce tit-for-tat fairness. But, seeds can represent an
important part of a \p2p session, see
Table~\ref{table_torrent_charac}. For this reason, it is fundamental
to have a notion of fairness that takes into account seeds.

In the following, we present two fairness criteria that take into
account the characteristics of leechers and seeds and the notion of
excess capacity:
\begin{itemize}
\item Any leecher $i$ with an upload speed $U_i$ should get a lower
  download speed than any other leecher $j$ with an upload speed $U_j>U_i$.
\item A seed should give the same service time to each leecher.
\end{itemize}

With these two simple criteria, leechers are allowed to use the
excess capacity, but not at the expense of leechers with a higher
level of contribution. Reciprocation is fostered and free riders are
penalized.  Seeds do not make a distinction between contributing
leechers and free riders. However, free riders cannot compromise the
stability of the system because the more there are contributing
leechers, the less the free riders receive from the seeds.

Tit-for-tat fairness can be extended to evenly distribute the capacity
of seeds to peers in a torrent. With this extension, tit-for-tat
fairness will verify our two fairness criteria. However, in the
context of peers with asymmetric capacity, finding a threshold that
maximizes the capacity of the system is a hard task that is not yet
solved in the context of a distributed system. Moreover, using a
default threshold may lead to a high inefficiency of the system.  We
will see in the following that the choke algorithm verifies our two
fairness criteria with a simple distributed algorithm that does not
require the complex computation of a threshold.

To summarize the above discussion, tit-for-tat fairness is not
appropriate in the context of \p2p file replication protocols like
\btns. For this reason, we proposed two new criteria of fairness, one
for leechers and one for seeds. It is beyond the scope of this study
to perform a detailed discussion of the fairness issues for \p2p
protocols. Our intent is to give a good intuition on how a \p2p
protocol should behave in order to achieve a reasonable level of
fairness.

In the following, we show on real torrents that the choke algorithm in
leecher state fosters reciprocation, and that the new choke algorithm in
seed state gives the same service time to each leecher. We conclude
that the choke algorithm is fair according to our two new fairness
criteria.

\subsubsection{Leecher State}
\label{sec:leecher-state}

\begin{figure}
\centering
\includegraphics[width=3.1in]{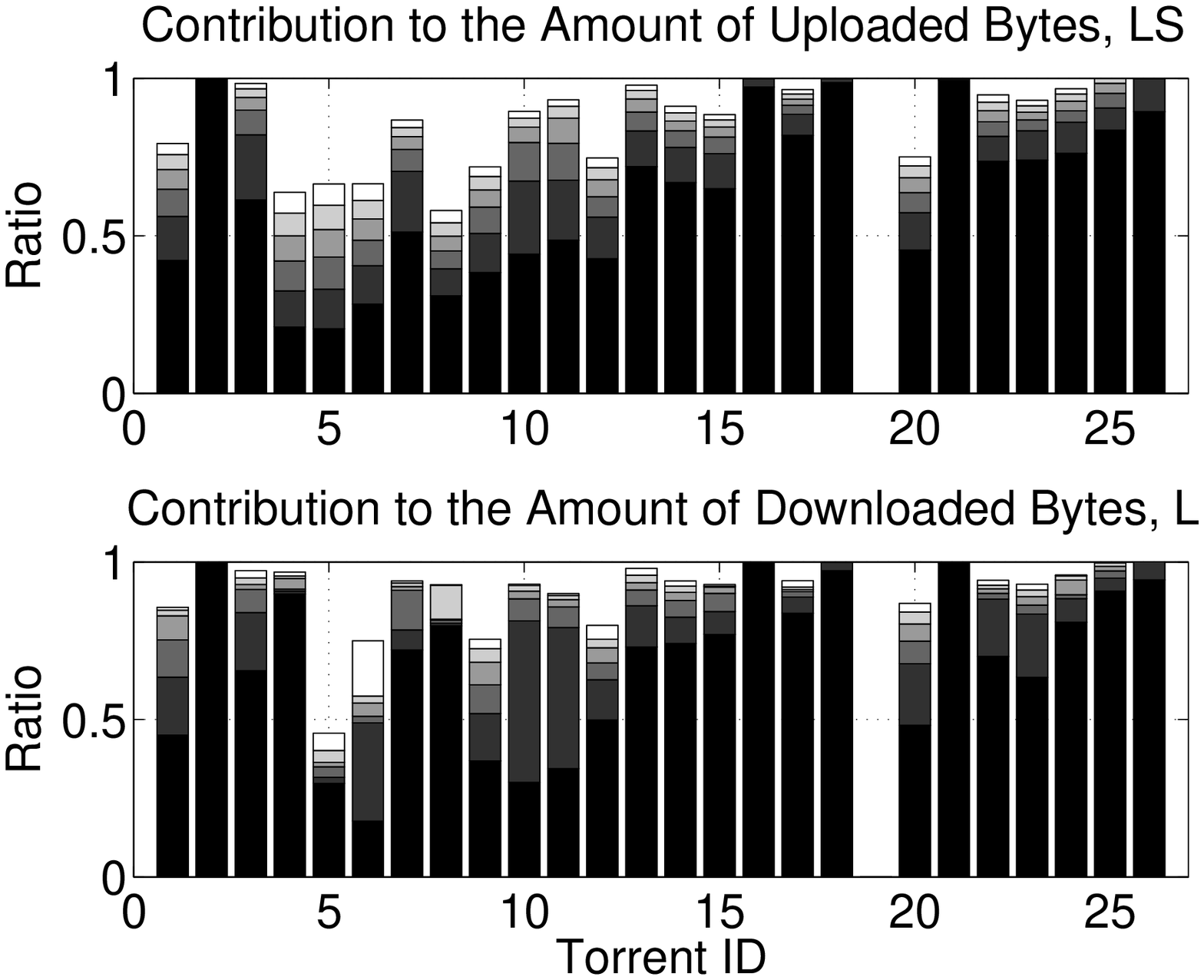}
\caption{\textmd{\textsl{Fairness characterization of the choke algorithm in leecher
  state for each torrent. \textbf{Top graph:} Amount of bytes uploaded from
  the local peer to remote peers. We created 6
sets of 5 remote peers each, the first set (in black) contains the 
5 remote peers that receive the most bytes from the local peer. Each
next set contains the next 5 remote peers. The
sets representation goes from black for the set containing the 5 best remote
downloaders, to white for the set containing the 25 to 30 best
downloaders.  \textbf{Bottom graph:} Amount of bytes downloaded
  from remote peers to the local peer. The same set construction is
  kept. Thus, this graph shows how much each set of downloaders, as
  defined in the top graph, uploaded to the local peer. 
}}}
\label{fig:all-ratio-ls}
\end{figure}

The choke algorithm in leecher state fosters reciprocation. We see in
Fig.~\ref{fig:all-ratio-ls} that peers that receive the most from the
local peer (top graph) are also peers from which the local peer
downloaded the most (bottom graph). Indeed, the same color in the
top and bottom graphs represents the same set of peers. All seeds are
removed from the data used for the bottom graph, as it is not possible
to reciprocate data to seeds. This way, a ratio of 1 in the bottom
graph represents the total amount of bytes downloaded from leechers.

Two torrents present a different characteristic. The local peer for
torrent 19 does not upload any byte in leecher state because due to the
small number of leechers in this torrent, the local peer in leecher
state had no leecher in its peer set.  Torrents 5, which is
in transient state, has a low level of reciprocation. This is
explained by a single leecher that gave to the local peer half of the
pieces, but who received few pieces from the local peer. The reason is
that this remote leecher was almost never interested in the local
peer. This problem is due to the low entropy of the torrent in
transient state.

Because the choke algorithm takes its decisions based on the current
download rate of the remote peers, it does not achieve a perfect
reciprocation of the amount of bytes downloaded and uploaded. However,
Fig.~\ref{fig:all-ratio-ls} shows that the peers from which the local
peer downloads the most are also the peers that receive the most
uploaded bytes. Thus there is a strong correlation between the amount
of bytes uploaded and the amount of bytes downloaded.

The above results show that with a simple distributed algorithm and
without any stringent reciprocation requirements, unlike tit-for-tat
fairness, one can achieve a good reciprocation. More importantly, the
choke algorithm in leecher state allows leechers to benefit from the
excess capacity. It is important to understand why the choke algorithm
achieves this good reciprocation. One reason is the way the active
peer set is built.  In the following, we focus on how the local peer
selects the remote peers to upload blocks to.

The choke algorithm in leecher state selects a small subset of peers
to upload blocks to. We see in Fig.~\ref{fig:all-ratio-ls}, top
graph, that the 5 peers that receive the most data from the local
peer (in black) represents a large part of the total amount of
uploaded bytes. At first sight, this behavior is expected from the
choke algorithm because a local peer selects the three fastest
downloading peers to upload to, see section~\ref{sec:choke-algorithm}.
However, there is no guarantee that these three peers will continue to
send data to the local peer. In the case they stop sending data to the
local peer, the local peer will also stop reciprocating to them.

\begin{figure}
\centering
\includegraphics[width=2.9in]{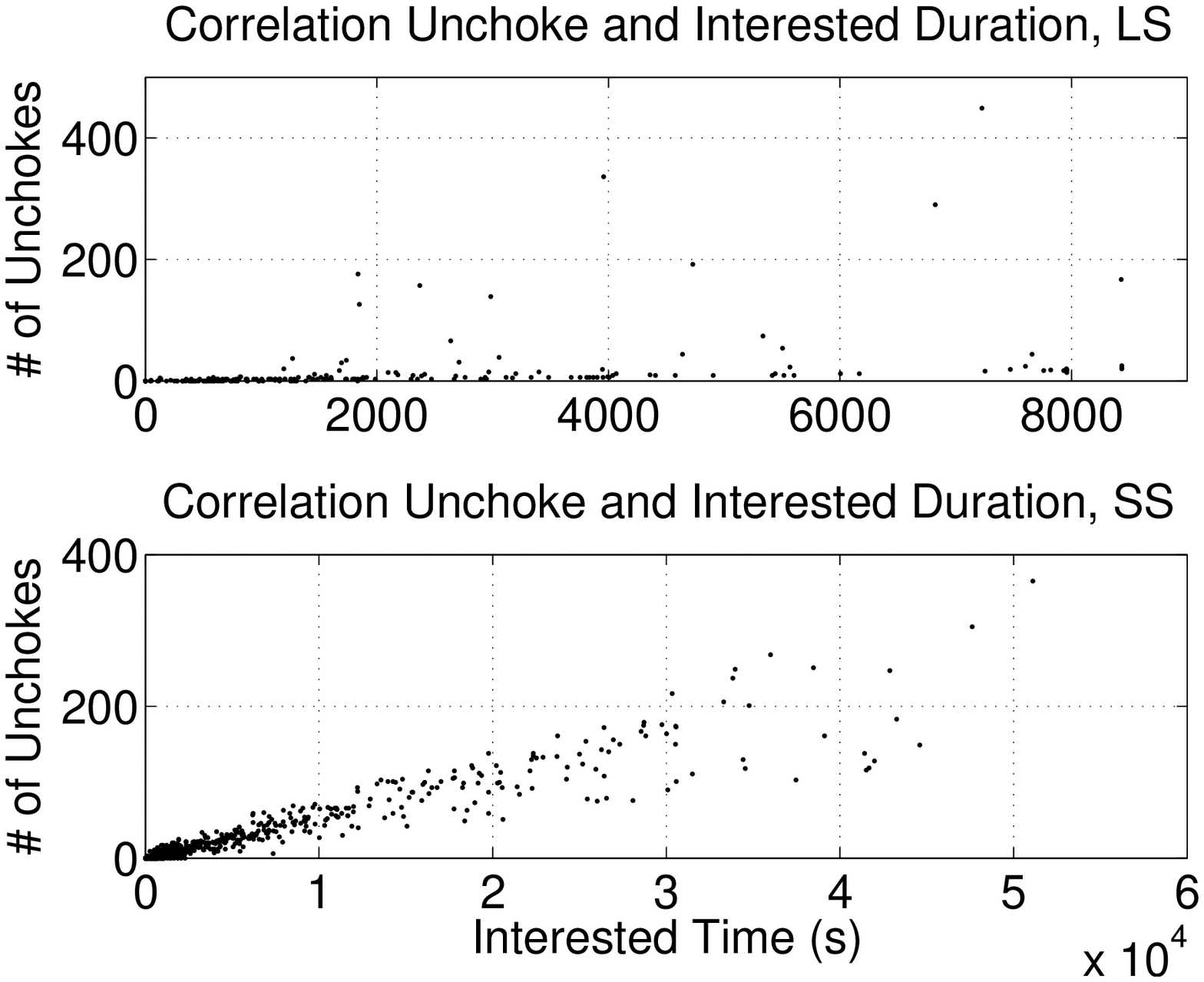}
\caption{\textmd{\textsl{Correlation between the number of unchokes
      and the interested time for each remote peer for torrent 7. For
      each remote peer, a dot represents the correlation between the
      number of times this remote peer is unchoked by the local peer
      and the time this remote peer is interested in the local peer.
      \textbf{Top graph:} Correlation when the local peer is in
      leecher state. \textbf{Bottom graph:} Correlation when the
      local peer is in seed state.}}}
\label{fig:corr-ls-ss}
\end{figure}

We focus on torrent 7 in order to understand
how this subset of peers is selected. Fig.~\ref{fig:corr-ls-ss} (top
graph) shows that most of the peers are unchoked few times and few
peers are unchoked frequently. The optimistic unchoke gives a
chance to each peer to be unchoked few times, whereas the regular
unchoke is used to unchoke frequently peers that send the fastest
to the local peer. The optimistic unchoke acts as a peer
discovery mechanism. The peers that are not unchoked at all are
either initial seeds, or peers that do not stay in the peer set long enough to
be optimistically unchoked. 

We see in  Fig.~\ref{fig:corr-ls-ss} (top graph) that there is no
correlation between the number of times a peer is unchoked and how long
a peer is interested in the local peer. However, we see that
the number of unchokes for the peers that are unchoked few times
increases slightly with the interested time duration. This is
because the optimistic unchoke takes at random a peer to be
optimistically unchoked. Thus the longer a peer is interested in the
local peer, the more likely it has to be optimistically unchoked.

Fig.~\ref{fig:all-ratio-ls} shows that for four torrents in transient
state, torrents 4, 5, 6 and 8, the amount of bytes uploaded by the 30
best remote peers is lower than for the other torrents. Torrents in
transient state have low entropy. Therefore, the peers are no longer
selected based only on their reciprocation level, but also on the
pieces available. For this reason, a larger set of peers receives
pieces from the local peer. Thus, a lower fraction of bytes uploaded
to the best remote peers.

In summary, we have seen that the choke algorithm fosters
reciprocation. One important reason is that each peer elects a small
subset of peers to upload data to. This stability improves the level
of reciprocation. We have seen that this stability is not due to a
lack of interest. Our guess is that the choke algorithm leads to an
equilibrium in the peer selection. The exploration of this equilibrium
is fundamental to the understanding of the choke algorithm efficiency.
It is beyond the scope of this study to do this analysis, but it is an
important area of future research.

\subsubsection{Seed State}
\label{sec:seed-state}

\begin{figure}
\centering
\includegraphics[width=2.9in]{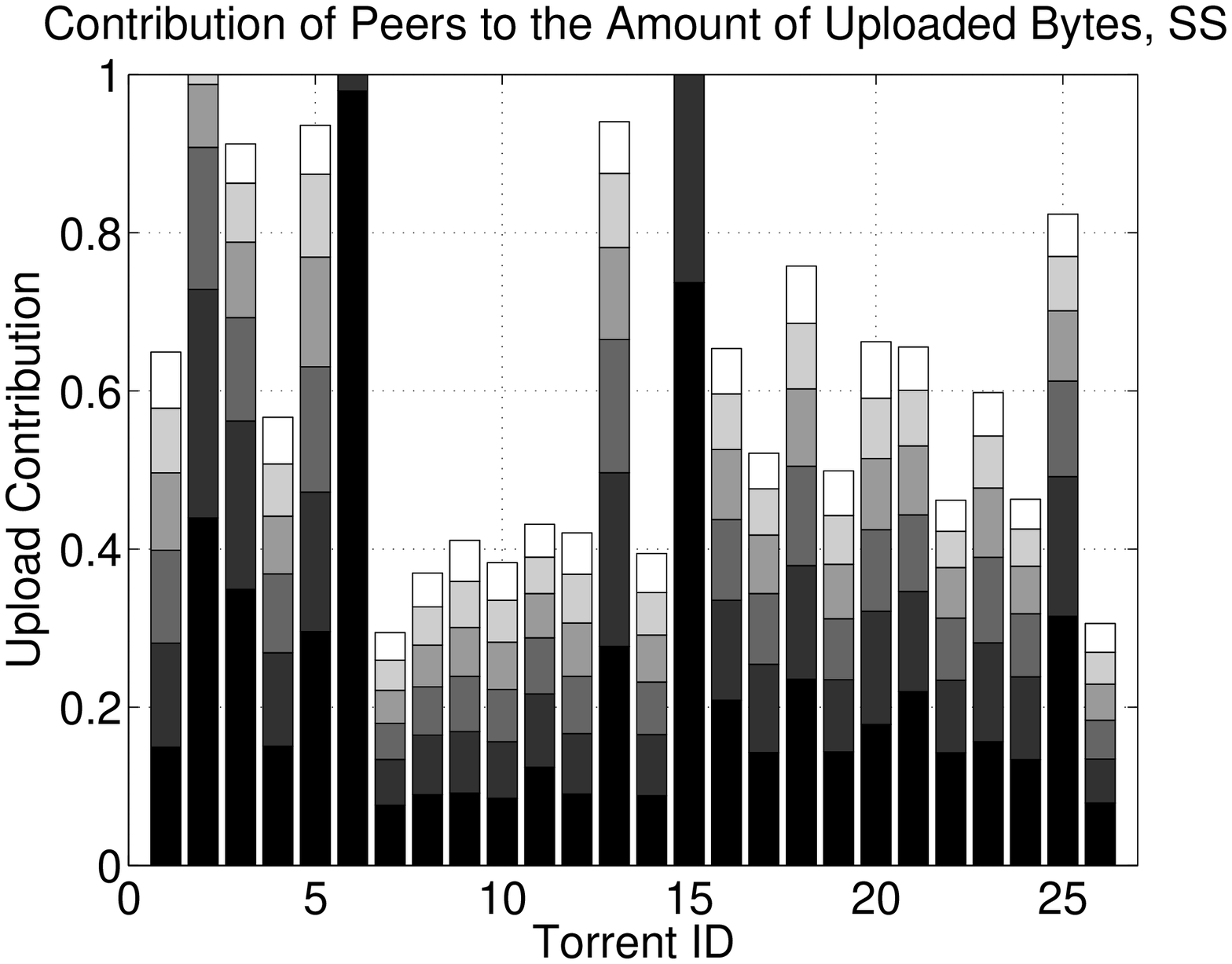}
\caption{\textmd{\textsl{Fairness characterization of the choke algorithm in seed
  state for each torrent. \textbf{Legend:} We created 6
  sets of 5 remote peers each, the first set (in black) contains the 5
  remote peers that receive the most bytes from the local peer. Each
  next set contains the next 5 remote peers. The set representation
  goes from black from the set containing the 5 best remote
  downloaders, to white for the set containing the 25 to 30 best
  downloaders.  
}}}
\label{fig:all-ratio-ss}
\end{figure}

The new choke algorithm in seed state gives the same service time to
each remote peer. We see in Fig.~\ref{fig:all-ratio-ss} that each peer
receives roughly the same amount of bytes from the local peer. The
differences among the peers are due to the time remote peers are
interested in the local peer. The more a remote peer is interested in
the local peer, the more times this remote
peer is unchoked. This is confirmed by Fig.~\ref{fig:corr-ls-ss}
(bottom graph) that shows a strong correlation between the time a
peer is interested in the local peer and the number of times the local
peer unchokes it.  For torrents 6 and 15 the five best downloaders
receive most of the bytes, because for both torrents there were less
than 10 remote peers that received bytes from the local peer.

This new version of the choke algorithm in seed state is the only one
to give the same service time to each leecher. This has three
fundamental benefits compared to the old version. First, as each
leecher receives a small and equivalent service time from the seeds,
the entropy of the pieces is improved. In contrast, with the old choke
algorithm, a few fast leechers can receive most of the pieces, which
decreases the diversity of the pieces. Second, free riders cannot
receive more than contributing leechers. In contrast, with the old
choke algorithm, a fast free rider can monopolize a seed. Third, the
resilience in transient phase is improved. Indeed, the initial
seed does not favor any leecher. Thus, if a leecher leaves the peer set,
it will only remove a small subset of the pieces from the torrent. In
contrast, with the old choke algorithm, the initial seed can send
most of the pieces to a single leecher. If this leecher leaves the
torrent, that will adversely impact the torrent and increase the time
in transient state.

In summary, the new choke algorithm in seed state gives the same
service to time to each leecher. This new algorithm is a significant
improvement over the old one. In particular, whereas the old choke
algorithm can be unfair and sensible to free riders, the new choke
algorithm is fair and robust to free riders.

\section{Related Work}
\label{sec:related-work}
Whereas \bt can be considered as one of the most successful \p2p
protocol, there are few studies on it.

Several analytical studies of {\btns}-like protocols exist
\cite{qiu04, yang04, biersack04}. Whereas they provide 
a good insight into the behavior of such protocols, the assumption of
global knowledge limits the scope of their conclusions.
Biersack et al. \cite{biersack04} propose an analysis of three content
distribution models: a linear chain, a tree, and a forest of
trees. They discuss the impact of the number of chunks
(what we call pieces) and of the number of simultaneous
uploads (what we call the active peer set) for each model. They show
that the number of chunks should be large and that the number of
simultaneous uploads should be between 3 and 5. 
Yang et al. \cite{yang04} study the service capacity of
{\btns}-like protocols. They show that the service capacity increases
exponentially at the beginning of the torrent and then scale well with
the number of peers. They also present traces obtained
from a tracker. Such traces are very different from ours, as they do
not allow to study the dynamics of a peer. Both studies presented in \cite{biersack04}
and \cite{yang04} are orthogonal to ours as they do not consider the
dynamics induced by the choke and rarest first algorithms.   
Qiu and Srikant \cite{qiu04} extend the initial
work presented in \cite{yang04} by providing an analytical solution to
a fluid model of \btns. Their results show the high
efficiency in terms of system capacity utilization of \btns, both in a
steady state and in a transient regime. Furthermore, the authors
concentrate on a game-theoretical analysis of the choke and
rarest first algorithms.  However, a major 
limitation of this analytical model is the assumption of global
knowledge of all peers to make the peer selection. Indeed, in a real
system, each peer has only a limited view of the other peers, which is
defined by its peer set. As a consequence, a peer cannot find the best
suited peers to send data to in all the peers in the torrent (global
optimization assumption), but in its own peer set (local and
distributed optimization). Also, the authors
do not evaluate the rarest first algorithm, but assume a uniform
distribution of pieces. Our study is complementary, as it provides
an experimental evaluation of algorithms with limited knowledge. In
particular, we show that the efficiency on real torrents is close to
the one predicted by the models. 

Felber et al. \cite{felber04} compare different peer and piece
selection strategies in static scenarios using simulations.  Bharambe
et al. \cite{bharambeApr06} present a simulation-based study of \bt
using a discrete-event simulator that supports up to 5000 peers. The
authors concentrate on the evaluation of the \bt performance by
looking at the upload capacity of the nodes and at the fairness
defined in terms of the volume of data served by each node. They
varied various parameters of the simulation as the peer set and active
peer set size. They provide important insights into the behavior of
\btns. However, they do not evaluate a peer set larger than 15 peers,
whereas the real implementation of \bt has a default value of 80
peers. This restriction may have an important impact on the behavior
of the protocol as the piece selection strategy is impacted by the
peer set size. The validation of a simulator is always hard to
perform, and the simulator restrictions may biased the results. Our
study provides real word results that can be used to validate
simulated scenarios. Moreover, our study is different because we do
not modify the default parameters of \btns, but we observed its
default behavior on a large variety of real torrents. Finally, we
provide new insights into the rarest first piece selection and on the
choke algorithm peer selection.  In particular, we argue that the
choke algorithm in its latest version is fair.

Pouwelse et al. \cite{pouwelse05} study the file popularity, file
availability, download performance, content lifetime and pollution
level on a popular \bt tracker site. This work is orthogonal to ours
as they do not study the core algorithms of \btns, but rather focus on
the contents distributed using \bt and on the users behavior.  The
work that is the most closely related to our study was done by Izal et
al. \cite{izal04}. In this paper, the authors provide seminal insights
into \bt based on data collected from a \textit{tracker} log for a
\textit{single} yet popular torrent, even if a sketch of a local
vision from a local peer perspective is presented. Their results
provide information on peers behavior, and show a correlation between
uploaded and downloaded amount of data. Our work differs from
\cite{izal04} in that we provide a thorough measurement-based analysis
of the rarest first and choke algorithms. We also study a large
variety of torrents, which allows us not to be biased toward a
particular type of torrent. Moreover, without pretending to answer all
possible questions that arise from a simple yet powerful protocol as
\btns, we provide new insights into the rarest first and choke
algorithms.

\section{Discussion}
\label{sec:discussion}
In this paper we go beyond the common wisdom that \bt performs well.
We have performed a detailed experimental evaluation of the rarest
first and choke algorithms on real torrents with varying
characteristics in terms of number of leechers, number of seeds, and
content sizes. Whereas we do not pretend to have reached completeness, our
evaluation gives a reasonable understanding of the behavior of both
algorithms on a large variety of real cases. 

Our main results are the following.
\begin{itemize}
\item The rarest first algorithm guarantees a close to ideal entropy
  on the presented torrents. In particular, it prevents the
  reappearance of rare pieces and of the last pieces problem.
\item We have found that torrents in a startup phase can have low
  entropy. The duration of this phase depends only on the
  upload capacity of the source of the content. In particular, the
  rarest first algorithm is not responsible of the low entropy during this phase.
\item The fairness achieved with a bit level tit-for-tat strategy is
  not appropriate in the context of \p2p file replication. We have
  proposed  two new fairness criteria in this context.
\item The choke algorithm is fair, fosters reciprocation, and is
  robust to free riders in its latest version.
\end{itemize}

Our main contribution is to show that on real torrents the rarest
first and choke algorithms are enough to have an efficient and viable
file replication protocol in the Internet. In particular, we discussed the benefits of
the new choke algorithm in seed state. This new algorithm outperforms
the old one and should replace it. We also identified two new
areas of improvement: the downloading speed of the first blocks, and
the duration of the transient phase. 

The rarest first algorithm is simple. It does not require global
knowledge or important computational resources. Yet, it guarantees a
peer availability, for the peer selection, close to the ideal one. We
do not see any striking argument in favor of a more complex solution
in the evaluated context. 

We do not claim that the choke algorithm is optimal. The understanding
of its equilibrium is an area of future research. However, it achieves a
reasonable level of efficiency, and most importantly it guarantees a viable
system by fostering reciprocation, preventing free riders to attack
the stability of the system, and using the excess capacity. 
Solutions based on a bit level tit-for-tat are not appropriate. 

Our conclusions only hold in the context we explored, i.e., \p2p file
replication in the Internet.
There are many different contexts where \p2p file replication can be
used: small files, small group of peers, dynamic groups in ad-hoc
networks, peers with partial connectivity, etc. All these contexts are
beyond the scope of this paper, but are interesting areas for future
research.

We also identified two areas of improvement. The time to deliver the
first blocks of data should be reduced. In the case of large contents, this
delivery time will marginally increase the overall download time. But,
in the case of small contents, the penalty is significant. Also, the
duration of the transient phase should be minimized as the low
entropy may results in a performance penalty. The way to solve these
problems is beyond the scope of this study, but is an interesting area
of future research. 

We believe that this work sheds a new light on two new algorithms that
enrich previous content distribution techniques in the Internet. \bt
is the only existing peer-to-peer replication protocol that exploits
these two promising algorithms in order to improve system capacity
utilization. We deem that the understanding of these two algorithms is
of fundamental importance for the design of future peer-to-peer
content distribution applications. 

\section*{Acknowledgment}
We would like to thank the anonymous reviewers, and also Chadi Barakat, Ernst W. Biersack, Walid
Dabbous,  Katia Obraczka, Thierry Turletti for their valuable comments.

\end{document}